\documentclass[12pt]{article}

\usepackage{graphicx}
\usepackage{amsmath}
\usepackage{amsfonts}
\numberwithin{equation}{section}

\def\normalorder{{\!\! \tiny\begin{array}{cc}\circ \\ \circ \\ \end{array}\!\!}} 
\def\thickone{{\rm 1\mskip-4.5mu l}}
\renewcommand{\frac}{\tfrac}

\begin{document}

\title{{\bf 
	The Lorentzian Space-Times of the\\ Orientation-Orbifold String Systems
}}

\author{
	M.B.Halpern\footnote{halpern@physics.berkeley.edu}\\
	Department of Physics\\
	University of California\\
	Berkeley, Ca. 94708, USA
}

\maketitle

\begin{abstract}

\noindent 
To illustrate our recent discussions of the target space-times in general
orbifold-string theories of permutation-type, we return here to a 
detailed analysis of some simple examples of this type, namely an explicit set of 
orientation-orbifold string systems. 
These orientation-orbifold string systems provide twisted, multisector generalizations of ordinary critical
open-closed bosonic string systems -- each such system exhibiting a unique 
graviton. Furthermore, each sector  $\sigma$ of each of these string systems shows the following properties: a) 26 effective degrees of freedom, 
b) a Lorentzian space-time with space-time dimension $D(\sigma)\leq 26$,  c) an $SO(D(\sigma)-1,1)$-invariant ordinary string subsystem 
with quantized intercept less than or equal one, and
d) an extra set of $(26 - D(\sigma))$ twisted fields which are $SO(D(\sigma)-1,1)$ scalars.
Subexamples of non-tachyonic strings and four-dimensional strings are noted. 
Additionally, we discuss certain subsets of physical states of these theories, concluding that these
investigations are so far consistent with the no-ghost conjecture for all
the Lorentzian orbifold-string theories of permutation-type.

\end{abstract}


\newpage

\tableofcontents

\newpage

\section{Introduction}

\noindent
The orbifold-string theories of permutation-type [1-8] are multisector 
candidates for \emph{new physical string systems}. 
The expectation [1] that these theories should be physical is based simply on  the principles
of orbifold theory and the fact that sets of copies of ordinary critical strings
are themselves physical. Any ``surprising'' consistencies encountered in the exploration
of these theories -- and we shall see a number of these below -- should be viewed in this perspective.

Using the techniques of the orbifold program [9-23], we have so far 
studied only the bosonic prototypes of these theories,  including 
\begin{subequations}
\begin{equation}
	\frac{U(1)^{26K}}{H_+},\,\, [\frac{U(1)^{26K}}{H_+}]_\text{open}, \quad H_+ \subset H(\text{perm})_K \times H'_{26}
\end{equation}
\begin{equation}
	\frac{U(1)^{26}}{H_-} = \frac{U(1)^{26}_L \times U(1)^{26}_R}{H_-},  \quad H_- \subset \mathbb{Z}_2(w.s.)\times H'_{26}
\end{equation}
\end{subequations}
where $U(1)^{26}$ is the critical closed string, $H(\text{perm})'_K$ 
permutes the copies of the closed string and $H'_{26}$ is any 
automorphism group of (the left-and right-movers of) the closed string.
The systems in Eq.~(1.1b) with $H(\text{perm})_2 = \mathbb{Z}_2(w.s.)$ are 
called the orientation-orbifold string theories [20,21,1,3,4,7,8], to which we shall return below.

As large examples of these systems, we have studied the automorphism groups [3,8]
\begin{equation}
	H'_{26} \subset (\pm \thickone)_d \times H(\text{perm})_{26-d}, \quad 1\leq d\leq 26
\end{equation} 
for which the divisors $H_\pm$ in Eq.~(1.1) then involve the \emph{two} permutation 
groups $H(\text{perm})_K$ and $H(\text{perm})'_{26-d}$. Both permutation groups
have been systematically analyzed [8] in terms of the respective cycle 
lengths $f_j(\sigma)$ (in $H(\text{perm})_K)$  and $F_J(\sigma)$ (in $H(\text{perm})'_{26-d}$) of their elements
\begin{subequations}
\begin{equation}
	\bar{\hat j} = 0, 1,\dots, f_j(\sigma)-1,\quad j=0,1,\dots, N(\sigma)-1
\end{equation}
\begin{equation}
	\bar{\hat J} = 0, 1,\dots, F_J(\sigma)-1,\quad J=0,1,\dots, N(\sigma)'-1
\end{equation}
\end{subequations}
in each twisted sector $\sigma$ of the orbifold-string theory.

So far we have presented discussions of the following topics:
\begin{itemize}
\item the extended actions and new twisted world-sheet permutation 
gravities [1,6]
\item BRST quantization [2,6] of each cycle in the twisted sectors
\item the extended physical-state conditions [3-8] at cycle central charge $\hat c_j(\sigma) = 26 f_j(\sigma)$
\item the cycle dynamics [3,7,8] of each sector, including the explicit form
of the extended Virasoro generators and twisted current algebras of each cycle.
\item the equivalent, reduced formulation [3,5,7,8] at reduced cycle central 
charge $c_j(\sigma)=26$ 
\item interacting examples [4,5]
\item target space-times and their symmetries [7,8].
\end{itemize}
In these investigations we have seen that, although the original 
formulation of cycle $j$ (at cycle central charge $\hat c_j(\sigma) = 26 
f_j(\sigma)$) is local [1], the equivalent, reduced formulation (at 
reduced cycle central charge $c_j(\sigma)=26$ is often simpler. In any 
case the two formulations can be straightforwardly reconstructed, each 
from the other, via a precise 1-1 map which generalizes (the inverse of) the
orbifold-induction procedure [9,3,7,8]. In particular, the \emph{target space-time dimension} 
of cycle $j$ in sector $\sigma$
\begin{equation}
	\hat D_j(\sigma) = D_j(\sigma)
\end{equation}
is invariant under the map. Moreover, these target space-times can be 
either Lorentzian or Euclidean [8].

Drawing on this general analysis, our task here is to focus on some simple examples
in further detail. In particular, we concentrate on the following large example  of 
\emph{orientation-orbifold string theories} [1,3,4,7,8]
\begin{subequations}
\begin{equation}
	\frac{U(1)^{26}}{H_-} = \frac{U(1)^{26}_L \times U(1)^{26}_R}{H_-},  
\end{equation}
\begin{equation}
H_{-} \subset \{\tau_{0} \times (\thickone)_{(d)} \times  H(\text{perm})'_{26-d}; 
\,\,\tau_{-} \times (\pm \thickone)_{(d)} \times H(\text{perm})'_{26-d}\}
\end{equation}
\begin{equation}
	1\leq d \leq 26
\end{equation}
\end{subequations}
where $\tau_{0}=1$ and $\tau_{-}$ is the non-trivial element of the 
world-sheet orientation-reversing $\mathbb{Z}_2$ called $\mathbb{Z}_2(w.s.)$.
These systems include [4] and generalize the ordinary critical open-closed string 
system , and consist of an equal number of twisted open- and closed-string sectors:
\begin{subequations}
\begin{equation}
	\text{open:}\quad \tau_{-},\quad f_j(\sigma) = 2,\quad \hat c_j(\sigma) = 52,\quad c_j(\sigma) =26
\end{equation}
\begin{equation}
	\text{closed:}\quad \tau_{0}=1,\quad f_j(\sigma) = 1,\quad \hat c_j(\sigma) = c_j(\sigma) = 26.
\end{equation}
\end{subequations}
The open-string sectors appear after the semicolon in Eq.~(1.5b), where 
one may choose the d-dimensional automorphism as either 
$(+\thickone)_{(d)}$ or $(-\thickone)_{(d)}$ but not both.
The closed-string sectors appear before the semicolon, and together form the ordinary space-time orbifold
\begin{equation}
\frac{U(1)^{26}}{H'_{26}},\quad H'_{26} \subset (\thickone)_{(d)} \times H(\text{perm})'_{26-d}
\end{equation}
where we have suppressed the division  by the trivial element of 
$\mathbb{Z}_2(w.s.)$. The unique graviton of each orientation-orbifold 
string system is found in the closed-string sector corresponding to the trivial
element of $H(\text{perm})'_{26-d}$.
The bulk of this paper focuses on the twisted open-string sectors, but we 
will return in Secs.~14 and 15 to assemble the full orientation-orbifold string theories.

The systems (1.5) were partially analyzed in Ref.~[3], and it is now 
known [8] that all sectors of all these orientation-orbifold string 
systems live on sector-dependent \emph{Lorentzian} target space-times, with 
space-time symmetry $SO(D_j(\sigma)-1,1)$. Here, we will work this out in 
detail, concluding that each sector of these theories contains an 
ordinary $D_{j}(\sigma)$-dimensional string subsector, plus a set of $(26-D_j(\sigma))$ extra
$SO(D_j(\sigma)-1,1)$-invariant twisted scalar fields. This gives a total of  
$c_{j}(\sigma)=26$ effective degrees of freedom in each sector, as in the ordinary 
open-closed string system.

When full computational details are required, we illustrate with the cyclic subset 
\begin{subequations}
\begin{equation}
H(\text{perm})'_{26-d} = \mathbb{Z}_{26-d}
\end{equation}
\begin{equation}
	1\leq d \leq 26.
\end{equation}
\end{subequations}
In these cases we compute the number of target space-time 
dimensions explicitly (see the Tables of Sec.~7), concluding that 
\begin{equation}
	\hat D_j(\sigma) = D_j(\sigma)\leq 26.
\end{equation}
Secs.~8 and 9 emphasize in particular the non-tachyonic strings  and the four-dimensional strings contained
in this subset of theories. 
 
The upper bound (1.9) is in fact a necessary  condition for the new string theories to be 
physical and, indeed, such gratifying consistency  -- as we have 
explained above -- is expected on general grounds for all $H(\text{perm})'_{26-d}$.
Arguments are given in Secs.~4,7 and 15 that the bound indeed holds 
for general $H(\text{perm})'_{26-d}$ , and  
slightly stronger conditions on the open-string space-time dimensionalities  are obtained 
in Eq.~(7.6).     

These results lead us to examine  a number of subsets of physical states 
(see Secs.~10,11 and 13) in the large example (1.5), concluding that so 
far these investigations are consistent with the no-ghost 
conjecture (see Ref.~[1] and Sec.~12)
for all Lorentzian orbifold-string theories of permutation-type. 
Sec.~13 notes in particular the ordinary $(D_{j}(\sigma)\leq 
26)$-dimensional open strings which live as subsectors of these theories
at quantized intercept less than or equal to one.

\section{The Twisted Open-String sectors at $\hat c = 52$}

\noindent The open-string sectors of the orientation-orbifold string 
systems (1.5)
\begin{subequations}
\begin{equation}
\{\text{open}\} = \{\tau_{-} \times (\pm \thickone)_{(d)} \times H(\text{perm})'_{26-d}\}
\end{equation}
\begin{equation}
	1\leq d \leq 26
\end{equation}
\end{subequations}
are associated with the non-trivial element $\tau_{-}$ of $\mathbb{Z}_2(w.s.)$. 
This element consists of a single j-cycle of length 2
\begin{equation}
	f_j(\sigma) = 2, \quad \hat c_j(\sigma) = 52, \quad j =0,\quad \bar{\hat j} = 0,1
\end{equation}
but we will for simplicity suppress the single cycle label $j=0$. 

Then we may write the extended physical-state conditions and orbifold 
Virasoro generators of open-string sector $\sigma$ as follows [3,8]:
\begin{subequations}
\begin{equation}
 (\hat L_{\hat j}((m+\frac{\hat j}{2})\geq 0) - \frac{17}{8} \delta_{m+\frac{\hat j}{2},0}) | \chi(\sigma)\rangle = 0,\quad \bar{\hat j} = 0,1
\end{equation}
\begin{equation}
	\begin{split}
 [\hat L_{\hat j}(m+\frac{\hat j}{2}), \hat L_{\hat \ell}(n + \frac{\hat\ell}{{2}})] = & (m-n + \frac{\hat j - \hat\ell}{2})  \hat L_{\hat j+\hat\ell}(m + n + \frac{\hat j + \hat\ell}{2}) + \\
    & + \frac{52}{12} (m + \frac{\hat j}{2})((m+\frac{\hat j}{2})^{2}-1) \delta_{m+n+\frac{\hat j+\hat\ell}{2},0}
  \end{split}
\end{equation}
\begin{equation}
   \begin{split}
   & \hat L_{\hat j}(m+\frac{\hat j}{2}) =  \hat\Delta_0(\sigma) \delta_{m+\frac{\hat j}{2},0} + 
          \frac{1}{4} G^{ab}_{(d)} \sum^1_{\hat\ell=0}\sum_{p\in \mathbb{Z}} 
             \normalorder \hat J_{\epsilon a\hat\ell} (p+ \frac{\epsilon + \hat\ell}{2}) \hat J_{-\epsilon,b,\hat j - \hat\ell}(m - p + \frac{\hat j -\hat \ell - \epsilon}{2}) \normalorder_M + \\
   & + \frac{1}{4} \sum_L \frac{1}{F_L(\sigma)} \sum^1_{\hat\ell=0} \sum_{\hat L=0}^{F_L(\sigma)-1} \sum_{p\in\mathbb{Z}}
		\normalorder \hat J_{\hat L L \hat \ell}(p + \frac{\hat L}{F_L(\sigma)} + \frac{\hat\ell}{2})
          \hat J_{-\hat L,L,\hat j-\hat\ell}(m -p - \frac{\hat L}{F_L(\sigma)} + \frac{\hat j - \hat\ell -\epsilon}{2})\normalorder_M
	\end{split}
\end{equation}
\begin{equation}
	\hat\Delta_0(\sigma) = \frac{13}{8} + \frac{1}{2} \hat{\delta}_0(\sigma)
\end{equation}
\begin{equation}
	\hat \delta_0(\sigma) = \hat\delta_0(\sigma)^{(26-d)} = 
        \frac{1}{4} \sum_L \sum_{\hat L=0}^{F_L(\sigma)-1} (\frac{2 \hat L}{F_L(\sigma)}-1)(2\theta(\frac{2\hat L}{F_L(\sigma)}\geq 1)-\frac{2\hat L}{F_L(\sigma)})\geq 0
\end{equation}
\begin{equation}
  a=0,1,\dots, d-1, \quad G_{(d)} =
   \left(
   \begin{array}{cc}
      -1 & 0 \\
      0  & \thickone
   \end{array}\right)_{(d)}
  , \quad \sum_L F_L(\sigma) = 26-d, \quad F_L(\sigma)\geq 1.
\end{equation}
\end{subequations}
\noindent We remind that the choice of either $\epsilon = 0$ or $1$ 
(but not both) corresponds to the choice of d-dimensional
automorphism $\omega_{(d)}=(\thickone)_{(d)}$ or $(-\thickone)_{(d)}$ 
respectively, and that neither choice makes any contribution $(\hat \delta_0(\sigma)^{(d)} = 
0)$ to the conformal-weight shifts $\hat\delta_0(\sigma)$.
The explicit form (2.3e) of the conformal-weight shift of sector $\sigma$ is therefore determined entirely
by the choice $\omega(\sigma)\in H(\text{perm})'_{26-d}$, which
is characterized here by its L-cycle lengths $\{F_L(\sigma)\}$.
This result was given in Ref. [3] in the notation $\bar{\hat j} = \bar u 
=0,1$ (with small letters for the $L$-cycles), and can also be obtained 
in the present notation as the special case $f_j(\sigma)=2$ in Ref.~ [8]. 

In these simple cases, the orbifold Virasoro generators $\{\hat L_{\hat 
j}\}$ of sector $\sigma$ in Eq.~(2.3c) satisfy the orbifold Virasoro 
algebra [9,17, 2-8] of order two in Eq.~(2.3b). The orbifold Virasoro 
algebra has an integral Virasoro 
subalgebra at $\hat c(\sigma)=52$
defined by the generators $\{\hat L_0(m)\}$.
The conventional mode normal-ordering indicated here
\begin{equation}
 \normalorder \hat A(\xi) \hat B(\eta) \normalorder_M \equiv \theta(\xi\geq 0) \hat B(\eta)\hat A(\xi) + \theta(\xi<0) \hat A(\xi)\hat B(\eta)
\end{equation}
will be employed throughout this paper.

In what follows, we often use the nomenclature of Ref.~[8], refering to 
the two kinds of twisted currents in Eq.~(2.3c) as those 
of \emph{type $(d)$} $(a=0,\dots,d-1)$ and those of \emph{type $(26-d)$}.
The rest of the algebra of these currents is easily read from Refs.~[3] or [8]
\begin{subequations}
\begin{equation}
	[\hat L_{\hat j}(m + \frac{\hat j}{2}), \hat J_{\epsilon a\hat \ell}(n + \frac{\epsilon +\hat\ell}{2})] = 
	-(n + \frac{\epsilon+\hat\ell}{2}) \hat J_{\epsilon a,\hat j+\hat\ell} (m+n+\frac{\epsilon + \hat j + \hat \ell}{2})
\end{equation}
\begin{equation}
	[\hat L_{\hat j}(m + \frac{\hat j}{2}), \hat J_{\hat L L\hat\ell}(n+\frac{\hat L}{F_L(\sigma)} + \frac{\hat\ell}{2})] = 
	-(n + \frac{\hat L}{F_L(\sigma)} + \frac{\hat\ell}{2}) \hat J_{\hat L L, \hat j + \hat\ell} (m + n + \frac{\hat L}{F_L(\sigma)} + \frac{\hat j + \hat \ell}{2}) 
\end{equation}
\begin{equation}
	[\hat J_{\epsilon a \hat j}(m+\frac{\epsilon+\hat j}{2}), 
	 \hat J_{\epsilon' b\hat\ell}(n + \frac{\epsilon'+\hat\ell}{2})]
	= 2 (m+\frac{\epsilon+\hat j}{2}) G^{(d)}_{ab} \delta_{\epsilon+\epsilon',0\text{ mod }2}
	\delta_{m+n+\frac{\epsilon+\epsilon'+\hat j+\hat \ell}{2},0}
\end{equation}
\begin{multline}
	[\hat J_{\hat J J \hat j}( m + \frac{\hat J}{F_J(\sigma)} + \frac{\hat j}{2}),
	 \hat J_{\hat L L \hat \ell}( n + \frac{\hat L}{F_L(\sigma)} + \frac{\hat \ell}{2})]=\\
	= \delta_{JL} 2 F_J(\sigma) (m+\frac{\hat J}{F_J(\sigma)} + \frac{\hat j}{2})
	 \delta_{\hat J+\hat L, 0\text{ mod }F_J(\sigma)} \delta_{M+N+\frac{\hat J+\hat L}{F_J(\sigma)} + \frac{\hat j + \hat \ell}{2},0}
\end{multline}	 
\begin{equation}
1 \leq d \leq 26,\,\, a,b=0,1,\, \dots d-1,\, \hat L=0,1,\,\dots F_L(\sigma)-1,\, L = 0,1,\dots N(\sigma)'-1
\end{equation} 
\end{subequations}
where $N(\sigma)'$ is the number of L-cycles in the element of $H(\text{perm})'_{26-d}$. Note that
the range of the parameter $d$ cannot be extended
to $d=0$ because $H(\text{perm})'_{26}$ is not an automorphism group
of the original Minkowski metric $G_{(26)}$ of $U(1)^{26}$.

For computational purposes, we will also need the periodicity conditions 
and the adjoints of all these operators [8]:
\begin{subequations}
\begin{equation}
	\hat L_{\hat j\pm 2}(m + \frac{\hat j\pm 2}{2}) = \hat L_{\hat j}(m \pm 1 + \frac{\hat j}{2})
\end{equation}
\begin{equation}
	\hat J_{\epsilon a, \hat j\pm 2} (m+ \frac{\epsilon + \hat j \pm 2}{2}) = 
	\hat J_{\epsilon a \hat j}(m\pm 1 + \frac{\epsilon+\hat j}{2})
\end{equation}
\begin{equation}
	\hat J_{-\epsilon, a, \hat j}(m + \frac{\hat j-\epsilon}{2}) = 
	\hat J_{\epsilon a \hat j   }(m - \epsilon + \frac{\epsilon + \hat j}{2})
\end{equation}
\begin{equation}
	\hat J_{\hat L\pm F_L(\sigma), L\hat\ell} (m + \frac{\hat L\pm F_L(\sigma)}{F_L(\sigma)}+\frac{\hat\ell}{2}) =
	\hat J_{\hat L L, \hat\ell\pm 2}(m + \frac{\hat L}{F_L(\sigma)} + \frac{\hat\ell \pm 2}{2}) =
	\hat J_{\hat L L \hat\ell}(m \pm 1 +  \frac{\hat L}{F_L(\sigma)} + \frac{\hat\ell}{2})
\end{equation}
\begin{equation}
	\hat L_{\hat j}(m + \frac{\hat j}{2})^\dagger = \hat L_{-\hat j}(-m-\frac{\hat j}{2})
\end{equation}
\begin{equation}
	\hat J_{\epsilon a \hat j}(m + \frac{\epsilon + \hat j}{2})^\dagger = 
	\hat J_{-\epsilon, a, -\hat j}(-m - \frac{(\epsilon + \hat j)}{2})
\end{equation}
\begin{equation}
	\hat J_{\hat L L \hat\ell} (m+ \frac{\hat L}{F_L(\sigma)} + \frac{\hat\ell}{2})^\dagger =
	\hat J_{-\hat L, L, -\hat\ell}(-m -\frac{\hat L}{F_L(\sigma)} - \frac{\hat\ell}{2}).
\end{equation}
\end{subequations}
It follows in particular that the generators of the Virasoro subalgebra 
satisfy the conventional generalized hermiticity $\hat L_0(m)^\dagger = \hat L_0(-m)$.

In each sector $\sigma$, a useful decomposition of the zero mode of the 
Virasoro generators is
\begin{equation}
	\hat L_0(0) = \hat L_0(0)^\dagger = \frac{1}{4} (-\hat P^2(\sigma) + \hat R(\sigma)) + \hat\Delta_0(\sigma)
\end{equation}
where $\hat R(\sigma)$ is the generalized number operator [3,8] of sector 
$\sigma$ and
\begin{subequations}
\begin{multline}
	\hat P^2(\sigma) = \hat P^2(\sigma)^\dagger = \eta^{ab}_{(d)} J_{\epsilon a\epsilon}(0) J_{-\epsilon, b, -\epsilon}(0)\\
	- \sum_L\frac{1}{F_L(\sigma)} (\hat J_{0L0}(\sigma) J_{0L0}(\sigma) + 
	\hat J_{\frac{F_L(\sigma)}{2}, L, 1}(0) \hat J_{-\frac{F_L(\sigma)}{2}, L, -1}(0))
\end{multline}
\begin{equation}
	\eta_{(d)} = -G_{(d)} = 
	   \left(
   \begin{array}{cc}
      1 & 0 \\
      0  & -\thickone
   \end{array}\right)_{(d)}
\end{equation}
\end{subequations}
is the \emph{momentum-squared operator} of the sector.
The last term of this expression contributes only when an L-cycle length $F_L(\sigma)$ is even.

We comment further on the set of zero modes $\{ \hat J(0)_\sigma\}$ or 
\emph{momenta} of sector $\sigma$,
which commute with each other and all other current modes of the sector
\begin{equation}
	[\{ \hat J(0)_\sigma\}, \hat J(\{\text{all})\} ] = 0.
\end{equation}
The number of zero modes then defines the \emph{target space-time 
dimension} [7,8] of sector $\sigma$, and for the open-string sectors of the orientation-orbifolds
we easily count
\begin{equation}
	\hat D(\sigma) \equiv \text{dim}\{ \hat J(0)_\sigma \} = 
	 d + N_O(\sigma)' + 2 N_E(\sigma)' 
\end{equation}
\begin{equation}
	N(\sigma)'= N_O(\sigma)' + N_E(\sigma)'
\end{equation}
where $N_{O,E}(\sigma)'$ are respectively the number of $L$-cycles of 
odd and even length $F_L(\sigma)$ in
$\omega(\sigma) \in H(\text{perm})'_{26-d}$. This result is implicit in  Ref.~[3] and
in agreement with the special case $f_j(\sigma) = 2$ in Ref.~[8].

The target space-time dimension (2.10) can also be expressed in the notation
\begin{equation}
	\hat D(\sigma) = d + \sum_L\alpha(L), \quad
	\alpha(L) \equiv \left\{
		\begin{array}{ll}

		1 & \text{for } F_L(\sigma) \text{ odd}  \geq 1 \\

		2 & \text{for } F_L(\sigma) \text{ even} \geq 2

		\end{array}
	\right. 
\end{equation}
where we have introduced the $L$-cycle function $\alpha(L)$. This function is natural in the orientation-orbifold
string systems, and appears as well in the following simple expression for the 
conformal-weight shifts of sector $\sigma$
\begin{subequations}
\begin{equation}
	\hat\delta_0(\sigma) = \frac{1}{24} \sum_L \{ \theta(F_L(\sigma) =\text{odd} \geq 1) 
		\frac{F_L^2(\sigma) - 1}{F_L(\sigma)}
		+ \theta(F_L(\sigma) = \text{even} \geq 2) \frac{F_L(\sigma) -4}{F_L(\sigma)} \}
\end{equation}
\begin{equation}
	= \frac{1}{24} (26-d - \sum_L \frac{\alpha^2(L)}{F_L(\sigma)}) , \quad\quad 
	\sum_L F_L(\sigma) = 26-d
\end{equation}
\end{subequations}
after performing the sum over $\hat L$ in Eq.~(2.3e). Note that
the first form of the conformal-weight shift in Eq.~(2.13a) shows 
explicitly that $\hat\delta_0(\sigma)\geq 0$, and the double inequalities
\begin{subequations}
\begin{equation}
	0\leq \hat\delta_0(\sigma) \leq \frac{1}{24}(26-d)
\end{equation}
\begin{equation}
	0\leq \sum_L \frac{\alpha^2(L)}{F_L(\sigma)} \leq 26-d, \quad \sum_L F_L(\sigma) = 26-d,\quad 1\leq d\leq 26
\end{equation}
\end{subequations}
then follow on comparison of the two forms in Eq.~(2.13).

The decomposition (2.7) and the extended physical-state condition (2.3a) 
also lead to the following result for the
\emph{ground-state momentum-squared} $\hat P^2(\sigma)_{(0)}$ of sector $\sigma$ [3,8]
\begin{subequations}
\begin{equation}
	\hat J_{\epsilon a \hat \ell} ((m+\frac{\epsilon+\hat\ell}{2})>0) 
		| 0, \hat J(0)_\sigma \rangle = 0
\end{equation}
\begin{equation}
	\hat J_{\hat L L\hat\ell} ((m+ \frac{\hat L}{F_L(\sigma)} + \frac{\hat \ell}{2}) > 0)
		| 0, \hat J(0)_\sigma \rangle = 0
\end{equation}
\begin{equation}
	\hat R(\sigma) | 0, \hat J(0)_\sigma\rangle = 0
\end{equation}
\begin{equation}
	\hat P^2(\sigma) | 0, \hat J(0)_\sigma\rangle = 
		\hat P^2(\sigma)_{(0)} | 0, \hat J(0)_\sigma\rangle
\end{equation}
\begin{equation}
	\hat P^2(\sigma)_{(0)} = 2 (\hat \delta_0(\sigma) - 1) \geq -2
\end{equation}
\end{subequations}
where $\hat \delta_0(\sigma)$ are the conformal-weight shifts and $\{ | 0, \hat J(0)_\sigma\rangle\}$ are the
oscillator-free eigenstates of the momenta $\{\hat J(0)_\sigma\}$. At 
this value of the momentum- squared, the ground-state can 
also be called the momentum-boosted twist-field state of sector $\sigma$.  The 
following simple form of the ground-state momentum-squared in terms of 
the L-cycle function
\begin{equation}
	\hat P^2(\sigma)_{(0)} = -\frac{1}{12} (d-2 + \sum_L \frac{\alpha^2(L)}{F_L(\sigma)}),
	\quad \sum_L F_L(\sigma) = 26-d
\end{equation}
is then obtained from the explicit form (2.13b) of the conformal-weight shifts.
Formulae are also known [3,8] for the \emph{level-spacing}
\begin{equation}
	\Delta(\hat P^2(\sigma) ) = 
	\Delta(\hat R^2(\sigma) ) =
	\left\{
		\begin{array}{ll}
		4 | m + \frac{\epsilon+\hat\ell}{2}| & \text{for } J((m+\frac{\epsilon+\hat\ell}{2})<0) \\
		4 | m + \frac{\hat L}{F_L(\sigma)} + \frac{\hat\ell}{2}| & \text{for } J((m+\frac{\hat L}{F_L(\sigma)} + \frac{\hat\ell}{2}) < 0)
		\end{array}
	\right.
\end{equation}
due to the addition of negatively-moded currents of either type to a 
lower-level state.

Using the adjoint operations in Eq.~(2.6) and the current algebras in 
Eq.~(2.5), we finally comment on the norms of the basis-states of sector 
$\sigma$. For the one-current states, we easily compute:
\begin{subequations}
\begin{equation}
	|| \hat J_{\epsilon a\hat\ell} ( (m+ \frac{\epsilon + \hat\ell}{2})<0) | 0, \hat J(0)_\sigma\rangle ||^2 
	 =  2 G^{(d)}_{aa} | m + \frac{\epsilon + \hat\ell}{2} | \,\, || \,\, | 0,\hat J(0)_\sigma\rangle \, ||^2
\end{equation}
\begin{equation}
	|| \hat J_{\hat L L \hat\ell} ((m+ \frac{\hat L}{F_L(\sigma)} + \frac{\hat\ell}{2})<0) | 0,\hat J(0)_\sigma\rangle ||^2
	= 2 F_L(\sigma) | m + \frac{\hat L}{F_L(\sigma)} + \frac{\hat\ell}{2} | \,\, || \,\,
		|0, \hat J(0)_\sigma \rangle ||^2.
\end{equation}
\end{subequations}
The norms of higher basis-states are similarly computed, including the 
strictly positive  norms of the general basis-state formed from the ``extra''
currents of type $(26-d)$:
\begin{multline}
	|| \prod_{m \hat L L\hat\ell} [ \hat J_{\hat L L \hat\ell} ((m+ \frac{\hat L}{F_L(\sigma)} +
	 \frac{\hat\ell}{2}) < 0)^{N_{m \hat L L \hat\ell}} ] |0, \hat J(0)_\sigma\rangle ||^2  = \\
	 = \prod_{m \hat L L\hat\ell} (N_{m \hat L L \hat\ell})!\,
		| F_L(\sigma) (2 m +\hat \ell) + 2 \hat L|^{N_{m \hat L L \hat\ell}}\,
			|| \,\, |0, \hat J(0)_\sigma\rangle ||^2.
\end{multline}
It is clear from these examples that, as in the ordinary open string, the 
\emph{only} negative-norm states in the basis  of sector $\sigma$ are 
those containing an odd number of \emph{time-like} modes $\{ \hat J_{\epsilon 0 \ell}\}$, that
is, $a=0$ of the currents of type $(d)$. This should come as no surprise, 
since the currents of type $(26-d)$ are the orbifoldization of copies  of
purely spatial currents in the untwisted sector $(U(1)^{26}_L \times U(1)^{26}_R)$ of the orbifold.

\section{The Equivalent Formulation at $c=26$}

\noindent It is now well understood [3-5,7,8] that the description above of the open-string
physical states at central charge $\hat c(\sigma)=52$ has an equivalent, 
reduced formulation at reduced central charge $c(\sigma)=26$.
Indeed, although we limit the discussion here to this case, we are 
in fact describing only the special case with j-cycle length 
$f_{j}(\sigma)=2$ included in the equivalent descriptions of the general cycle dynamics [7,8] 
at cycle central charge $\hat c_j(\sigma)=26 f_j(\sigma)$ and  uniform 
reduced cycle central charge $c_j(\sigma)=26$.

In the present case, the reduced formulation of sector $\sigma$ is 
obtained from the following 1-1 map [3]

\begin{subequations}
\begin{equation}
	L(M) \equiv 2 \hat L_{\hat\ell}(m + \frac{\hat\ell}{2}) - \frac{13}{4} \delta_{m + \frac{\hat\ell}{2},0}
\end{equation}
\begin{equation}
	J_{\epsilon a}(M + \epsilon) \equiv \hat J_{\epsilon a \hat\ell} ( m + \frac{\epsilon + \hat\ell}{2}), \quad a = 0,1,\dots, d-1
\end{equation}
\begin{equation}
	J_{\hat L L}(M + 2\frac{\hat L}{F_L(\sigma)}) \equiv \hat J_{\hat L L \hat \ell}( m + \frac{\hat L}{ F_L(\sigma)} + \frac{\hat\ell}{2})
\end{equation}
\begin{equation}
	M\equiv 2 m + \bar{\hat\ell}, \quad \bar{\hat\ell} = 0,1
\end{equation}
\end{subequations}
between the hatted operators at $\hat c(\sigma)=52$ and the unhatted operators
at $c(\sigma) = 26$. Note that $M\in \mathbb{Z}$ covers the integers once
for $\hat\ell$ in its fundamental range $\bar{\hat\ell}$. For the reduced 
currents of type (26-d) in Eq.~(3.1c), we emphasize that the factor 2 in 
the characteristic moding $\{2\hat{L}/F_{L}(\sigma)\}$ represents the effect 
on the element $\omega'(\sigma) \in H({\text perm})_{26-d}'$ due to the unwinding of the non-trivial 
element of $\mathbb{Z}_2(w.s.)$. In fact such 
maps, including the extension to arbitrary j-cycles [7,8], are only modestly-generalized versions of (the inverse of) the so-called
orbifold-induction procedure [9] -- which was originally used to construct cyclic
permutation orbifolds at an early stage of the orbifold program [10-23]. 

In terms of the unhatted operators, the map gives the following 
equivalent, reduced formulation of the  physical states of sector $\sigma$ at 
reduced central charge $c(\sigma)=26$:
\begin{subequations}
\begin{equation}
	( L(M \geq 0)) - \delta_{M,0} ) | \chi(\sigma) \rangle = 0
\end{equation}
\begin{equation}
	[L(M), L(N)] = (M-N) L(M + N) + \frac{26}{12} M (M^2 -1) \delta_{M+N, 0}
\end{equation}
\begin{multline}
	L(M) = \hat\delta_0(\sigma) \delta_{M,0} + \frac{1}{2} G_{(d)}^{ab} 
		\sum_{Q\in \mathbb{Z}} \normalorder J_{\epsilon a}(Q+\epsilon) J_{-\epsilon, b} (M-Q-\epsilon) \normalorder_M + \\
     + \frac{1}{2} \sum_L \frac{1}{F_L(\sigma)} 
				\sum_{\hat L=0}^{F_L(\sigma)-1} \sum_{Q\in \mathbb{Z}} 
					\normalorder J_{\hat L L}(Q + 2\frac{\hat L}{F_L(\sigma)} ) 
						J_{-\hat L,L} (M - Q - 2 \frac{\hat L}{F_L(\sigma)}) \normalorder_M
\end{multline}
\begin{equation}
	F_L(\sigma) \geq 1, \quad \sum_L F_L(\sigma) = 26-d, \quad 1\geq d \geq 26, \quad a = 0,1,\dots,d-1.
\end{equation}
\end{subequations}
The Minkowski-space metric $G_{(d)}$ and the conformal-weight
shifts $\hat \delta_0(\sigma)$ of open-string sector $\sigma$ are given in Eqs.~(2.3f)
and (2.3e), (2.13) respectively, and we note in particular that the extended 
physical-state conditions (2.3a) have been mapped into the
\emph{ordinary} physical-state condition (3.2a) -- with unit intercept -- for each sector $\sigma$.
It should also be emphasized that the states $\{|\chi(\sigma)\!>\}$ 
described in the reduced formulation 
are exactly the same physical states described at $\hat{c}(\sigma)=52$, now rewritten 
in terms of the reduced currents.
Finally, because the map preserves the sign of the mode numbers, the mode normal-ordering in the Virasoro generators (3.2c)
\begin{equation}
	\normalorder A(\xi) B(\eta) \normalorder_M = 
		\theta(\xi \geq 0) B(\eta) A(\xi) + \theta(\xi<0) A(\xi) B(\eta)
\end{equation}
is isomorphic to the mode normal-ordering defined 
at $\hat c=52$ in Eq. (2.4).

We remind that the two kinds of reduced currents here are called [8] those of type $(d)$ $(a=0,\dots,d-1)$ and those of type $(26-d)$.
The map straightforwardly 
completes the reduced formulation from the results of the previous 
section,  beginning with the algebra of the reduced currents: 
\begin{subequations}
\begin{equation}
	[L(M), J_{\epsilon a}(N+\epsilon)] = -(N+\epsilon) J_{\epsilon a}(M+N+\epsilon), \quad a=0,1,\dots,d-1
\end{equation}
\begin{equation}
	[L(M), J_{\hat L L} (N+\frac{2\hat L}{F_L(\sigma)})] =
		-(N +\frac{2\hat L}{F_L(\sigma)} ) J_{\hat L L} (M+N+\frac{2\hat L}{F_L(\sigma)})
\end{equation}
\begin{equation}
	[J_{\epsilon a}(M+\epsilon), J_{\epsilon' b}(N+\epsilon')] =
		G^{(d)}_{ab}(M+\epsilon) \delta_{\epsilon+\epsilon' , 0\text{ mod }2} \delta_{M+N+\frac{\epsilon + \epsilon'}{2},0}
\end{equation}
\begin{equation}
	[J_{\hat J J}(M+\frac{2\hat J}{F_J(\sigma)}), J_{\hat L L}(N + \frac{2\hat L}{F_L(\sigma)})] = 
		\delta_{JL} F_J(\sigma) (M + \frac{2 \hat L}{F_L(\sigma)} ) \delta_{\hat J + \hat L, 0\text{ mod } F_J(\sigma)} \delta_{M + N + 2 \frac{\hat J + \hat L}{F_J(\sigma)}, 0}
\end{equation}
\begin{equation}
	\hat L =0, 1, \dots, F_L(\sigma)-1, \quad L=0,1,\dots,N(\sigma)'-1, \quad \sum_L F_L(\sigma)=26-d.
\end{equation}
\end{subequations}
We also recall here that the values $\epsilon = 0$ or $1$ correspond to the original
automorphisms $(\omega)_d = (\thickone)_d$ or $(-\thickone)_d$ respectively.
As a simple application of the L-cycle data in Eq.~(3.4e), let us check explicitly that the number of 
degrees of freedom in the reduced description of each sector $\sigma$
\begin{equation}
	\sum_a + \sum_{L\hat L} = d + \sum_L F_L(\sigma) = 26
\end{equation}
agrees with the reduced central charge $c(\sigma) = 26$ in the Virasoro 
algebra (3.2b).

We give next the reduced form of the periodicity relations and the adjoint operations
\begin{subequations}
\begin{equation}
  J_{-\epsilon, a}(M-\epsilon) = J_{\epsilon a} ((M-2\epsilon) + \epsilon)
\end{equation}
\begin{equation}
  J_{\hat L \pm F_L(\sigma), L} (M+2 \frac{\hat L \pm F_L(\sigma)}{F_L(\sigma)}) = J_{\hat L L}(M \pm 2 + 2 \frac{\hat L}{F_L(\sigma)})
\end{equation}
\begin{equation}
  J_{\epsilon a}(M+\epsilon)^\dagger = J_{-\epsilon, a}(-M-\epsilon)
\end{equation}
\begin{equation}
  J_{\hat L L}(M+2 \frac{\hat L}{F_L(\sigma)})^\dagger = J_{-\hat L, L}(-M - 2\frac{\hat L}{F_L(\sigma)})
\end{equation}
\begin{equation}
  L(M)^\dagger = L(-M).
\end{equation}
\end{subequations}
where Eq.~(3.6e) is ordinary generalized hermiticity for the reduced 
Virasoro generators at $c(\sigma) = 26$.

The reduced zero modes at $c(\sigma) = 26$ are in 1-1 correspondence with 
the zero modes of the original formulation at $\hat c(\sigma) = 52$
\begin{subequations}
\begin{equation}
  J(0)_\sigma = \hat J(0)_\sigma
\end{equation}
\begin{equation}
  \text{dim}\{ J(0)_\sigma \} = \text{dim} \{ \hat J(0)_\sigma \}
\end{equation}
\end{subequations}
and we remark that the three types of reduced zero modes $\{ J_{\epsilon 
a}(0), J_{0L}(0) \text{ and } J_{F_L(\sigma)/2,L}\}$ 
occur respectively at $M = -\epsilon, 0$ and $1$. As above, the third type of 
zero mode exists only when the cycle-length $F_L(\sigma)$ is even, and so
we verify the invariance of the \emph{target space-time dimension}
\begin{equation}
  D(\sigma) = \hat D(\sigma) = d + N_O(\sigma)' + 2 N_E(\sigma)' =  d + \sum_L\alpha(L)
\end{equation}
under the reduction procedure (see Eqs.~(2.10) and (2.12)).

Similarly, we find the reduced momentum-squared operator
\begin{subequations}
\begin{equation}
  L(0) = \frac{1}{2} (-P^2(\sigma) + R(\sigma)) + \hat{\delta}_{0}(\sigma)
\end{equation}
\begin{equation}
  P^2(\sigma) = \hat P^2(\sigma) = \eta^{ab}_{(d)} \sum_{Q\in \mathbb{Z}} J_{\epsilon a}(0) J_{-\epsilon,b}(0)
      -\sum_L\{\tilde J_{0L}(0) \tilde J_{0L}(0) + \tilde J_{\frac{F_L(\sigma)}{2},L}(0) \hat J_{-\frac{F_L(\sigma)}{2},L}(0) \}
\end{equation}
\begin{equation}
  \tilde J_{0L}(0) \equiv \frac{1}{\sqrt{F_L(\sigma)}} J_{0L}(0), 
  \quad \tilde J_{\frac{F_L(\sigma)}{2},L}(0) \equiv \frac{1}{\sqrt{F_L(\sigma)}} J_{\frac{F_L(\sigma)}{2}, L}(0)  
\end{equation}
\end{subequations}
where the $d$-dimensional Minkowski metric $\eta_{(d)}$ is given in 
Eq.~(2.8b) and we have rescaled the last two terms (of type $(26-d)$) as 
shown in Eq.~(3.9c). 
All the operators $L(0)$, $P^2(\sigma)$ and $R(\sigma)=\hat{R}(\sigma)$ 
in this decomposition are 
hermitian,
and the explicit form of the generalized number operator $R(\sigma)$ is given in Refs.~[3,8].

We turn next to some simple properties of the physical states $\{ | 
\chi(\sigma) \rangle \}$, as defined by the reduced physical-state condition (3.20).
We remind that the physical states  are the \emph{same} 
as those defined by the extended physical-state conditions (2.3a). Indeed 
all states, including the physical states as well as the basis states, 
are inert under the map (3.1) -- which only relabels the operators. For 
good book-keeping however, we  should here imagine having used the map to replace  
the negatively-moded (hatted) currents of these states in the original formulation at 
$\hat c(\sigma) = 52$ by  the negatively-moded (unhatted) currents at $c(\sigma) = 26$.

More precisely, we have the \emph{same} ground state in each sector $\sigma$
\begin{subequations}
\begin{equation}
  | 0, J(0)_\sigma \rangle = |0, \hat J(0)_\sigma\rangle
\end{equation}
\begin{equation}
  J_{\epsilon a}((M+\epsilon)>0) |0, J(0)_\sigma \rangle = J_{\hat L L}(( n + 2\frac{\hat L}{F_L(\sigma)})>0 ) |0, J(0)_\sigma\rangle = 0
\end{equation}
\begin{equation}
  R(\sigma) | 0, J(0)_\sigma \rangle = 0
\end{equation}
\begin{equation}
  P^2(\sigma)_{(0)} = \hat P^2(\sigma)_{(0)} = 2 (\hat \delta_0(\sigma) - 1) 
    = -\frac{1}{12} (d-2 +\sum_L \frac{\alpha^{2}(L)}{F_L(\sigma)} ), \quad \sum_L F_L(\sigma) = 26-d
\end{equation}
\end{subequations}
with the \emph{same} ground-state momentum-squared in Eq.~(3.10d).
Including the commutators (3.4a,b), the level-spacings of the physical 
states on the addition of a negatively-moded current
\begin{equation}
  \Delta(P^2(\sigma)) = \Delta(R(\sigma)) = 
\left\{
		\begin{array}{ll}

		2 | M+\epsilon |  & \text{for } J_{\epsilon a}((M + \epsilon)<0) \\

		2 | M+\frac{2\hat L}{F_L(\sigma)} | & \text{for } J_{\hat L L}((M + \frac{2 \hat L}{F_L(\sigma)})<0)

		\end{array}
	\right. 
\end{equation}
are also obtained in the reduced formulation of sector $\sigma$. Recalling the
definition $M = 2 m+\bar{\hat \ell}$ in Eq.~(3.1d), it is easily checked 
that these level-spacings are the \emph{same} as those given in Eq.~(2.17) at $\hat c(\sigma)=52$.

The current algebras (3.4b,c) and the adjoint operations (3.6b,c) of the reduced 
formulation also allow the straightforward computation of the following
norms:
\begin{subequations}
\begin{equation}
  || J_{\epsilon a}((M+\epsilon)<0) |0, J(0)_\sigma \rangle ||^2 = 
    G_{aa}^{(d)} | M+\epsilon|\, || \, |0,J(0)_\sigma\rangle ||^2
\end{equation}
\begin{equation}
  || J_{\hat L L} ((M+\frac{2 \hat L}{F_L(\sigma)}) < 0) |0, J(0)_\sigma\rangle ||^2 =
    F_L(\sigma) | M + \frac{2 \hat L}{F_L(\sigma)} |\, || \,|0,J(0)_\sigma\rangle ||^2
\end{equation}
\begin{equation}
  || \prod_{Q \hat L L} J_{\hat L L}((Q + \frac{2 \hat L}{F_L(\sigma)}) < 0)^{N_{\hat L L}(Q)} |0, J(0)_\sigma\rangle ||^2
    = \prod_{Q\hat L L} N_{\hat L L}(Q)! |F_L(\sigma)Q + 2 \hat L|^{N_{\hat L L}(Q)} ||\, |0, J(0)_\sigma\rangle ||^2.
\end{equation}
\end{subequations}
Using again the definition $M=2 m+\bar{\hat \ell}$, we see that these 
norms are equal to those computed at $\hat c(\sigma)=52$ in Eqs.~(2.18),(2.19).
Indeed, because states are inert under the map, we understand that the inner products $\langle \alpha | \beta \rangle$ of
any two  states are invariant under the reduction procedure to $c(\sigma)=26$.

We will return to further analysis of the physical states at various 
points below, including in particular the discussion of Sections 10-12.


\section{The Enhanced Lorentz Symmetry $SO(D(\sigma)-1,1)$}

\noindent In string theory, the target space-time symmetry of the system is 
determined by the \emph {integer-moded sequences} [8] -- whose zero modes are 
the momenta of the string.
The number of integer-moded sequences in sector $\sigma$
\begin{equation}
  D(\sigma) = D(\sigma)^{(d)} + D(\sigma)^{(26-d)} = d+N_O(\sigma)' + 2 N_E(\sigma)'
\end{equation}
is of course the same as the dimension of the target space-time of the sector.

Let us first mention the contribution to this number from the 
integer-moded sequences of type $(d)$, which are contained entirely in the second term 
of the Virasoro generators (3.2c). This term can be expressed in the familiar form 
\begin{subequations}
\begin{equation}
  L(M)_{(d)} \equiv -\frac{1}{2} \eta_{(d)}^{ab} \sum_{Q\in\mathbb{Z}} \normalorder 
      J_a(Q) J_b(M-Q) \normalorder_M
\end{equation}
\begin{equation}
  \eta_{(d)} = -G_{(d)} = 
    \left( \begin{array}{cc}
      1 & 0 \\
      0 & -\thickone
    \end{array} \right)_{(d)}, \quad D(\sigma)^{(d)} = d
\end{equation}
\end{subequations}
after the shift $Q\to Q-\epsilon$ under the sum and the relabeling
\begin{equation}
  J_a(M)\equiv J_{\epsilon a}(M), \quad a=0,\dots, d-1
\end{equation}
for $\epsilon =0$ or $1$. These generators satisfy a Virasoro algebra by 
themselves,
with central charge $c=d$ and what we may call a ``preliminary'' Lorentz symmetry $SO(d-1,1)$.

But this is not the whole story in the new string theories, because these systems
also contain the integer-moded sequences of type $(26-d)$. We include 
these contributions as higher values of the index $a$ in the following definition:
\begin{subequations}
\begin{equation}
  J_a(M) \equiv \frac{1}{\sqrt{F_L(\sigma)}} \{ J_{0L}(M), J_{F_L(\sigma)/2,L}(M) \}, \quad a=d,\dots, D(\sigma)-1
\end{equation}
\begin{equation}
  D(\sigma)^{(26-d)} = N_O(\sigma)' + 2 N_E(\sigma)' =  \sum_L\alpha(L).
\end{equation}
\end{subequations}
The rescaling here is motivated by the form of the momentum-squared 
operator in Eq.~(3.9).

Then we may express the physical-state condition (3.2) of sector $\sigma$ in 
the following simple form
\begin{subequations}
\begin{equation}
  (L(M\geq 0)-\delta_{M,0}) |\chi(\sigma)\rangle = 0
\end{equation}
\begin{equation}
  [ L(M), L(N) ] = (M-N) L(M+N) +\frac{26}{12} M(M^2-1) \delta_{M+N,0}
\end{equation}
\begin{multline}
  L(M) = \delta_{M,0} \hat\delta_0(\sigma) - 
    \frac{1}{2} \eta^{ab}_{(D(\sigma))} \sum_{Q\in\mathbb{Z}} \normalorder J_a(Q) J_b(M-Q)\normalorder_M +\\
    + \frac{1}{2} \sum_L \frac{1}{F_L(\sigma)} \sum_{\hat L\neq 0, F_L(\sigma)/2}^{F_L(\sigma)-1} \sum_{Q\in\mathbb{Z}} \normalorder J_{\hat L L}(Q+\frac{2\hat L}{F_L(\sigma)}) J_{-\hat L,L}(M-Q-\frac{2\hat L}{F_L(\sigma)})\normalorder_M
\end{multline}
\begin{equation}
  L(0) = \frac{1}{2} (-P^2(\sigma) + R(\sigma)) + \hat\delta_0(\sigma)
\end{equation}
\begin{equation}
  P^2(\sigma) = \eta^{ab}_{(D(\sigma))} J_a(0) J_b(0)
\end{equation}
\begin{equation}
  \eta_{(D(\sigma))} =     \left( \begin{array}{cc}
      1 & 0 \\
      0 & -\thickone
    \end{array} \right)_{(D(\sigma))},\quad D(\sigma) = d+N_O(\sigma)' + 2 N_E(\sigma)'
\end{equation}
\end{subequations}
where we have collected \emph {all} the integer-moded sequences of the 
sector in the second 
term of the Virasoro generators. We remind that the explicit form of the conformal-weight shift $\hat\delta_0(\sigma)$ and the ground-state
mass-squared $P^2(\sigma)_{(0)}$ are given respectively in Eqs.~ (2.13) and (2.16).


Counting the extended Virasoro generators (2.3c) and the reduced Virasoro 
generators (3.2c), this is the third and -- as we shall see -- most transparent form of the 
dynamics of open-string sector $\sigma$. Before giving the physical
interpretation of this result however, we should re-express the remaining 
structure of the reduced system in our new notation.

After some algebra, we find for the currents in Eq.~(4.5c):

\begin{subequations}
\begin{equation}
  [L(M), J_a(N) ] = -N J_a(M+N), \quad a=0,1,\dots, D(\sigma)-1
\end{equation}
\begin{equation}
  [L(M), J_{\hat L L}(N+\frac{2\hat L}{F_L(\sigma)}) ] = -(N + \frac{2\hat L}{F_L(\sigma)}) J_{\hat L L}(M+N+\frac{2\hat L}{F_L(\sigma)})
\end{equation}
\begin{equation}
  [J_a(M), J_b(N)] = \eta_{ab}^{(D(\sigma))} N \delta_{M+N, 0}
\end{equation}
\begin{equation}
  [J_{\hat J J}(M+\frac{2\hat J}{F_J(\sigma)}) , J_{\hat L L}(N + \frac{2\hat L}{F_L(\sigma)})] = 
    \delta_{JL} F_J(\sigma) (M+\frac{2\hat L}{F_L(\sigma)}) \delta_{\hat J + \hat L,0\text{ mod }F_L(\sigma)} \delta_{M+N+\frac{2(\hat J+\hat L)}{F_L(\sigma)},0}
\end{equation}
\begin{equation}
  [\{ J(\text{int}) \}, \{ J(\text{frac}) \}] =0
\end{equation}
\begin{equation}
  J_a(M)^\dagger = J_a(-M), \quad J_{\hat L L}(M + \frac{2\hat L}{F_L(\sigma)})^\dagger = J_{-\hat L, L} (-M -\frac{2\hat L}{F_L(\sigma)}). 
\end{equation}
\end{subequations}
In this list, the currents shown are divided into the integer-moded
currents $\{ J(\text{int}) \}$ and the truly fractional-moded currents $\{ J(\text{frac}) \}$, with $\hat L\neq 0$ or $F_L(\sigma)/2$.
Similarly, the periodicity condition for $\{ J(\text{frac}) \}$ is the 
same as given in Eq.~(3.6b). We also find the level-spacing
\begin{equation}
  \Delta (P^2(\sigma)) = \Delta(R(\sigma)) = 
  \left\{
		\begin{array}{ll}

		2 |M|                               & \text{for } J_a(M<0),\, a=0,1,\dots, D(\sigma)-1 \\

		2 |M+\frac{2 \hat L}{F_L(\sigma)} | & \text{for } J_{\hat L L}((M +\frac{2 \hat L}{F_L(\sigma)})<0),\, \hat L\neq 0, \frac{F_L(\sigma)}{2}

		\end{array}
	\right. 
\end{equation}
in this notation.

We emphasize that the parameter $\epsilon = 0$ or $1$ (corresponding to $(\omega)_d = (\pm \thickone)_{(d)}$) 
no longer appears in this reduced formulation, owing to our redefinition 
(4.3), so the physical spectrum of each open-string sector is independent of $\epsilon$.

We are now able to state the \emph{enhanced or full Lorentz symmetry} [8] of 
open-string sector $\sigma$ in our large example of orientation-orbifold string systems.
It is clear from the current algebra (4.6c) and the corresponding adjoint operation (4.6f) that 
the second term of the Virasoro generators
\begin{subequations}
\begin{equation}
  L(M)_{(D(\sigma))} \equiv -\frac{1}{2} \eta_{(D(\sigma))}^{ab} \sum_{Q\in\mathbb{Z}} 
    \normalorder J_a(Q) J_b(M-Q) \normalorder_M
\end{equation}
\begin{equation}
  c_{(D(\sigma))} (\sigma) \equiv \sum_a\, = D(\sigma) = d + \sum_L\alpha(L)  =  d+\,N_O(\sigma)' + 2 N_E(\sigma)'
\end{equation}
\end{subequations}
is a set of \emph {ordinary} $SO(D(\sigma)-1, 1)$-invariant open-string 
Virasoro generators
on the $D(\sigma)$-dimensional Lorentzian target space-time. As in the 
ordinary string, each  
of the integer-moded
currents $\{ J_a(M), a=0,\dots,D(\sigma)-1 \}$ transforms as a $D(\sigma)$-dimensional
Lorentz vector under this Lorentz group.
 
The first and third terms of 
Eq.~(4.5c) form another (commuting) set of Virasoro generators $\{ 
L_\text{frac}(M) \}$ for the currents $\{ J(\text{frac}) 
\}$, which are  extra fractional-moded $SO(D(\sigma)-1,1)$-invariant 
scalar fields. . It is not surprising that the first and third terms of (4.5c) should form a 
Virasoro subsystem because integer-moded 
sequences (of either type) can not contribute [8] to the conformal-weight 
shifts (see Eq.~(2.13)). 
Using the sum rule in Eq.~(3.4e), we may compute the  central charge  
of $\{ L_\text{frac}(M) \}$  by counting as follows:
\begin{multline}
  c_\text{frac}(\sigma) \equiv \sum_L \sum_{\hat L\neq 0,\frac{F_L(\sigma)}{2}} =
    \sum_L (F_L(\sigma)-\alpha(L))\\ = 
      (26-d) - (N_O(\sigma)' + 2 N_E(\sigma)' ) = 26-D(\sigma).
\end{multline}
This summation tells us that there are exactly $(26-D(\sigma))$ extra fractionally-moded
$SO(D(\sigma)-1,1)$-invariant scalar fields $\{ J(\text{frac}) \}$ in the open-string system, 
which is consistent with the sum
\begin{equation} 
 c_{(D(\sigma))} (\sigma) + c_\text{frac}(\sigma)= c(\sigma)= 26.
\end{equation}
Moreover, since the number of fractionally-moded fields cannot be 
negative, the computation (4.9) provides a simple argument that the 
target-space dimensionality satisfies $D(\sigma) 
\leq 26$ for all $H(\text{perm})'_{26-d}$. We shall check this conclusion
more directly in Sec.~7, where somewhat stronger conditions are obtained 
for the open-string dimensionalities $D(\sigma)$.    
 
\newpage

We conclude this section with three simple examples of our development so far:

\noindent a) $d=26$, $H(\text{perm})'$ absent with $\sum_L \frac{\alpha^2(L)}{F_L(\sigma)} \equiv 0$

\noindent b) $d=25$, $H(\text{perm})'_1$ trivial with
    \begin{equation}
      F_L(\sigma) = N(\sigma)' = N_O(\sigma)' = 1
    \end{equation}
    
\noindent c) the trivial element $(\thickone)_{26-d} \in H(\text{perm})'_{26-d}$, $1\leq d \leq 26$ with
\begin{equation}
  L = 0,1, \dots, 25-d,\,\, \hat L=0,\,\, F_L(\sigma)=1,\,\, N(\sigma)' = N_O(\sigma)' = 26-d.
\end{equation}
In all three  cases we find the system
\begin{equation}
  L(M) = L(M)_{(26)} = - \frac{1}{2} \eta^{ab}_{(26)} \sum_{Q\in\mathbb{Z}} \normalorder J_a(Q) J_b(M-Q) \normalorder_M
\end{equation}
\begin{equation}
  D(\sigma) = 26, \quad \hat\delta_0(\sigma) = 0, \quad P^2(\sigma)_{(0)}=-2, \quad \Delta(P^2(\sigma)) = 2 |M| \text{ for } J_a(M<0)
\end{equation}
so that each of these examples is spectrally-equivalent to an ordinary $26$-dimensional $SO(25,1)$-invariant open string.


\section{Ground-State Lemmas}

We consider further the ground-state momentum-squared (3.10d) and the 
double inequality (2.14) for each sector $\sigma$ of any $H(\text{perm})'_{26-d}$
\begin{subequations}
\begin{equation}
  P^2(\sigma)_{(0)} = 2 (\hat\delta_0(\sigma)-1) = -\frac{1}{12} (d-2 + \sum_L \frac{\alpha^2(L)}{F_L(\sigma)})
\end{equation}
\begin{equation}
  F_L(\sigma) \geq 1, \quad \sum_L F_L(\sigma) = 26-d, \quad 0<\sum_L \frac{\alpha^2(L)}{F_L(\sigma)} \leq 26-d, \quad 1\leq d\leq 25.
\end{equation}
\end{subequations}
Combining these results we find the double inequality  for the 
ground-state momentum-squared of sector $\sigma$:
\begin{equation}
  -2\leq P^2(\sigma)_{(0)} < \frac{1}{12}(2-d), \quad 1\leq d\leq 25.
\end{equation}
The case $d=26$ (and hence $H(\text{perm})'$ absent) was discussed in the previous section.

Most of the ground states in our large example are therefore tachyonic
\begin{equation}
  -2\leq P^2(\sigma)_{(0)} <0, \quad 2\leq d\leq 25
\end{equation}
which is not surprising in these bosonic prototypes.

For $d=1$ (and hence $H(\text{perm})'_{25}$), the ground-state mass-squared 
takes the form
\begin{subequations}
\begin{equation}
  P^2(\sigma)_{(0)} = \frac{1}{12}(1-\sum_L \frac{\alpha^2(L)}{F_L(\sigma)}), \quad \sum_L F_L(\sigma) = 25
\end{equation}
\begin{equation}
  -2\leq P^2(\sigma)_{(0)} < \frac{1}{12}
\end{equation}
\end{subequations}
so these sectors are the only cases where non-tachyonic ground-states are possible.

\section{The Single-Cycle Sectors of $H(\mathrm{perm})'_{26-d} = \mathbb{Z}_{26-d}$}

\noindent As other simple examples of our results, let us compute explicitly for the 
single-cycle elements of the cyclic groups
\begin{equation}
  \omega(\sigma) \in H(\text{perm})'_{26-d} = \mathbb{Z}_{26-d}, \quad 1\leq d\leq 25
\end{equation}
\begin{equation}
  \rho(\sigma) = F_L(\sigma) = 26-d, \quad L=0, \quad N(\sigma)'=1
\end{equation}
where $\rho(\sigma)$ is the order of $\omega(\sigma)$.

Then we easily obtain for $d$ odd the $SO(d,1)$-symmetric sectors
\begin{subequations}
\begin{multline}
  L(M) = \hat\delta_0(\sigma)\delta_{M,0} - \frac{1}{2}\eta^{ab}_{(d+1)} 
    \sum_{Q\in\mathbb{Z}} \normalorder J_a(Q)J_b(M-Q)\normalorder_M + \\
    + \frac{1}{2(26-d)} \sum_{\hat L=1}^{25-d} \sum_{Q\in\mathbb{Z}}
      \normalorder J_{\hat L 0}(Q+\frac{2\hat L}{26-d}) J_{-\hat L,0}(M-Q-\frac{2\hat L}{26-d}) \normalorder_M
\end{multline}
\begin{equation}
  P^2(\sigma) = \eta^{ab}_{(d+1)} J_a(0)J_b(0), \quad D(\sigma)=d+1
\end{equation}
\begin{equation}
  \hat\delta_0(\sigma) =\frac{1}{24}(26-d-\frac{1}{26-d})
\end{equation}
\begin{equation}
  P^2(\sigma)_{(0)} = - \frac{1}{12} (d-2+\frac{1}{26-d}), \quad 1\leq d \text{ odd }\leq 25
\end{equation}
\end{subequations}
and for $d$ even the $SO(d+1,1)$-symmetric sectors
\begin{subequations}
\begin{multline}
  L(M) = \hat\delta_0(\sigma) \delta_{M,0} - \frac{1}{2} \eta^{ab}_{(d+2)} 
    \sum_{Q\in\mathbb{Z}} \normalorder J_a(Q) J_b(M-Q)\normalorder_M + \\
    + \frac{1}{2(26-d)} \sum_{\hat L\neq 0, \frac{26-d}{2}}^{25-d} \sum_{Q\in\mathbb{Z}}
      \normalorder J_{\hat L 0}(Q+\frac{2\hat L}{26-d}) J_{-\hat L,0}(M-Q-\frac{2\hat L}{26-d}) \normalorder_M
\end{multline}
\begin{equation}
  P^2(\sigma) = \eta^{ab}_{(d+2)} J_a(0)J_b(0), \quad D(\sigma) = d+2
\end{equation}
\begin{equation}
  \hat\delta_0(\sigma) = \frac{1}{24}(26-d -\frac{4}{26-d})
\end{equation}
\begin{equation}
  P^2(\sigma)_{(0)} = -\frac{1}{12} (d-2 + \frac{4}{26-d}),\quad 2\leq d\text{ even }\leq 24.
\end{equation}
\end{subequations}
The ground-state momentum-squared (6.3d) was given earlier for prime $d$ in Ref.~[3].

In agreement with the ground-state lemma (5.3), these single-cycle ground 
states are indeed tachyonic for all $2\leq d<25$. Among these, only the 
cases $d=24$
(the non-trivial element of $H(\text{perm})'_2 = \mathbb{Z}_2$ with $F_L(\sigma)=2$) and 
$d=25$ ($H(\text{perm})'_1 = \mathbb{Z}_1$ trivial with $F_L(\sigma)=1$) realize
the minimum value $P^2(\sigma)_{(0)}=-2$ of the unshifted tachyonic 
ground state, and it is easily checked that both of these cases
\begin{multline}
  d=24, 25: \quad L(M) = L(M)_{(26)} = -\frac{1}{2} \eta^{ab}_{(26)}
    \sum_{Q\in\mathbb{Z}} \normalorder J_a(Q)J_b(M-Q)\normalorder_M,\\
    \hat\delta_0(\sigma) = 0, \quad P^2(\sigma)_{(0)}=-2, \quad D(\sigma)=26, \quad \Delta(P^2(\sigma))=2|M|
\end{multline}
are spectrally equivalent to an ordinary $26$-dimensional 
$SO(25,1)$-invariant open string. 

Indeed, following the discussion of the previous section, the only non-tachyonic ground state
is found at $d=1$, which selects here the non-trivial single-cycle elements of $H(\text{perm})'_{25} = \mathbb{Z}_{25}$:
\begin{equation}
  d=1: \quad D(\sigma)=2, \quad P^2(\sigma)_{(0)}=\frac{2}{25} < \frac{1}{12}.
\end{equation}
In this $2$-dimensional $SO(1,1)$-invariant string, the automorphism 
$(\omega)_{d=1}= (\pm\thickone)_1$ contributes the time dimension while 
the element of  $\mathbb{Z}_{25}$ contributes the single spatial dimension.
Further discussion of the non-tachyonic sectors is found in Sec.~8.

The special cases noted in Eqs.~(6.5) and (6.6) are in fact the extrema 
of the single-cycle series, which satisfies more generally the following double inequalities:
\begin{equation}
  1\leq d \leq 25 : \quad 2\leq D(\sigma) \leq 26, \quad \frac{2}{25} \geq P^2(\sigma)_{(0)} \geq -2.
\end{equation}
Beyond the single cycle series, we shall see below that the minimal (unshifted) ground-state mass-squared $P^2(\sigma)_{(0)}=-2$ is always
associated to the maximal (ordinary) target space-time dimension $D(\sigma)=26$.

\section{The 24 Cyclic Groups and $D(\sigma)\leq 26$}

\noindent Following the evaluation of the single-cycle series in the previous 
section, we expand our inquiry here to include all sectors $\sigma$ of 
the 24 relevant cyclic groups:
\begin{subequations}
\begin{equation}
  \omega(\sigma) = e^{-2\pi i \frac{\sigma}{26-d}} \in H(\text{perm})'_{26-d} = \mathbb{Z}_{26-d}
\end{equation}
\begin{equation}
  1\leq d \leq 25, \quad \sigma = 0, 1, \dots, 26-d.
\end{equation}
\end{subequations}
For convenience we have included here the trivial group $\mathbb{Z}_1$ at $d=25$ and the trivial sector $\sigma=0$ of
each group, both of which have been described above (see Eqs.~(4.10) and (4.11)). 
The cyclic group $\mathbb{Z}_{26-d}$ has $25-d$ non-trivial elements (sectors) $\sigma = 1,\dots, 25-d$.
\newpage

The explicit form (7.1a) of the group elements is useful for counting the number
of sectors $\sigma$ of each cycle-type, and each cycle-type is easily 
described in our notation as follows
\begin{subequations}
\begin{equation}
  L = 0,1, \dots, \frac{26-d}{\rho(\sigma)}-1, \quad \hat L = 0, 1, \dots, \rho(\sigma) -1
\end{equation}
\begin{equation}
  F_L(\sigma) = \rho(\sigma), \quad \forall L
\end{equation}
\begin{equation}
  N(\sigma)' = \sum_L = \frac{26-d}{\rho(\sigma)}, \quad \sum_L F_L(\sigma) = \rho(\sigma) N(\sigma)' = 26-d
\end{equation}
\end{subequations}
where $\rho(\sigma)$ is the order of $\omega(\sigma) \in \mathbb{Z}_{26-d}$. 
The results below
record the detailed evaluation of the target space-time dimension of sector $\sigma$ [8]
\begin{equation}
  D(\sigma) = d+N_O(\sigma)' + 2 N_E(\sigma)', \quad N(\sigma)' = N_O(\sigma)' + N_E(\sigma)' 
\end{equation}
and hence the target space-time symmetry $SO(D(\sigma)-1,1)$ of the sector, where $N_{O,E}(\sigma)$ 
are respectively the number of cycles of odd and even length $F_L(\sigma)$ in $\omega(\sigma)$.
We shall also include some further remarks in this section on the ground-state
mass-squared $P^2(\sigma)_{(0)}$ in Eq.~(5.1a).

The result $D(\sigma)=26$ and $P^2(\sigma)_{(0)} = -2$ have already been 
recorded for the trivial sector $\omega(0)=1$ ($\sigma=0$ with $\rho(0)=1$) of each group, 
and this case is always spectrally-equivalent to an ordinary $26$-dimensional  $SO(25,1)$-invariant
open string. We shall therefore list below only the non-trivial 
sectors $\sigma = 1, \dots, 26-d$ of each group.

We will however encounter the space-time dimension $D(\sigma) =26$ a 
number of other times in the tables below, and we have checked that each 
occurence of this dimension is also equivalent to an ordinary 26-dimensional
open string with $P^2(\sigma)_{(0)} = -2$.

Let us illustrate our enumeration first with the simplest cases, namely 
the prime cyclic groups $\mathbb{Z}_{26-d}$ with $26-d$ prime. 
Each of these has $(25-d)$  non-trivial single-cycle sectors $\sigma$ 
with $L=0$ and length $F_{0}(\sigma) = 26-d$, shown in Table 1.

\newpage

\begin{table}[h]
  \begin{center}
  \begin{tabular}{c|c|c|c||c|c}
\hline\hline
$H(\text{perm})'$	&	$d$	&	$D(\sigma)$	&	symmetry	&	$D(\sigma)_c$	&	symmetry	\\
\hline
$(\mathbb{Z}_1\equiv S_1)$	&	25	&	26	&	$SO(25,1)$	&	26	&	$SO(25,1)$	\\ \hline
$\mathbb{Z}_2$	            &	24	&	26	&	$SO(25,1)$	&	25	&	$SO(24,1)$	\\ \hline
$\mathbb{Z}_3$	            &	23	&	24	&	$SO(23,1)$	&	24	&	$SO(23,1)$	\\ \hline
$\mathbb{Z}_5$            	&	21	&	22	&	$SO(21,1)$	&	22	&	$SO(21,1)$	\\ \hline
$\mathbb{Z}_7$	            &	19	&	20	&	$SO(19,1)$	&	20	&	$SO(19,1)$	\\ \hline
$\mathbb{Z}_{11}$	        &	15	&	16	&	$SO(15,1)$	&	16	&	$SO(15,1)$	\\ \hline
$\mathbb{Z}_{13}$	        &	13	&	14	&	$SO(13,1)$	&	14	&	$SO(13,1)$	\\ \hline
$\mathbb{Z}_{17}$	        &	9	&	10	&	$SO(9,1)$	&	10	&	$SO(9,1)$	\\ \hline
$\mathbb{Z}_{19}$	        &	7	&	8	&	$SO(7,1)$	&	8	&	$SO(7,1)$	\\ \hline
$\mathbb{Z}_{23}$	        &	3	&	4	&	$SO(3,1)$	&	4	&	$SO(3,1)$	\\
\hline\hline
  \end{tabular}
  \end{center}
  \caption{The prime cyclic groups}
\end{table}




Only the columns of this table before the double vertical line are relevant for the 
open-string sectors discussed here. (The last two columns of this and the 
following tables record the corresponding
data for the closed-string sectors, which we will explain in later 
sections of the paper.) 
Being single-cycle sectors, these cases are included in the results of the previous
section, and we remind here that all these ground states are tachyonic 
because $d\geq 2$. We notice also 
that $\mathbb{Z}_{23}$ provides our first 22 examples of $SO(3,1)$-invariant 
\emph{four-dimensional strings}, a subject to which we return in Sec.~9.

We continue our survey of the cyclic groups in the following sector notation:
\begin{subequations}
\begin{equation}
	n(\sigma)[F_L(\sigma)]^{N(\sigma)'}
\end{equation}
\begin{equation}
	N(\sigma)' = \frac{26-d}{F_L(\sigma)}, \quad L=0,1,\dots, \frac{26-d}{F_L(\sigma)}-1, \quad \sigma = 1,\dots, 25-d.
\end{equation}
\end{subequations}
Here $n(\sigma)$ is the number of sectors $\sigma $ with \emph{cycle-type} $[F_L]^{N'}$, that is
$N(\sigma)'$ cycles of length $F_L(\sigma)$. In the sector notation the 
non-trivial sectors of the prime cyclic groups above
read $(25-d)[26-d]^1$, and Table 2 gives a selection of non-prime cyclic 
groups in this notation.

\newpage

\begin{table}[h]
  \begin{center}
  \begin{tabular}{c|c|c|c|c||c|c}
\hline\hline
$H(\text{perm})'$	&	$d$	&	cycle-type & $D(\sigma)$	&	symmetry	 &	$D(\sigma)_c$	&	symmetry	\\ \hline

$\mathbb{Z}_4$	  &	22	&	$2[4]^1$ 	 & 24           & $SO(23,1)$ &	23	          &	$SO(22,1)$	\\
                  &	22	&	$1[2]^2$ 	 & 26           & $SO(25,1)$ &	24	          &	$SO(23,1)$	\\ \hline

$\mathbb{Z}_6$    & 20  & $2[6]^1$   & 22           & $SO(21,1)$ &  21            & $SO(20,1)$  \\
                  & 20  & $2[3]^2$   & 22           & $SO(21,1)$ &  22            & $SO(21,1)$  \\
                  & 20  & $1[2]^3$   & 26           & $SO(25,1)$ &  23            & $SO(22,1)$  \\ \hline

$\mathbb{Z}_8$    & 18  & $1[8]^1$   & 20           & $SO(19,1)$ &  19            & $SO(18,1)$  \\
                  & 18  & $5[4]^2$   & 22           & $SO(21,1)$ &  20            & $SO(19,1)$  \\
                  & 18  & $1[2]^4$   & 26           & $SO(25,1)$ &  22            & $SO(21,1)$  \\ \hline

$\mathbb{Z}_9$    & 17  & $6[9]^1$   & 18           & $SO(17,1)$ &  18            & $SO(17,1)$  \\
                  & 17  & $2[3]^3$   & 20           & $SO(19,1)$ &  20            & $SO(19,1)$  \\ \hline

$\vdots$          & $\vdots$ & $\vdots$  & $\vdots$  & $\vdots$ &  $\vdots$           & $\vdots$  \\ \hline

$\mathbb{Z}_{22}$ &  4  & $10[22]^1$ &  6           & $SO(5,1)$  &  5             & $SO(4,1)$  \\
                  &  4  & $10[11]^2$ &  6           & $SO(5,1)$  &  6             & $SO(5,1)$  \\
                  &  4  & $1[2]^{11}$  & 26           & $SO(25,1)$ &  15            & $SO(14,1)$  \\ \hline

$\mathbb{Z}_{24}$ &  2  & $17[24]^1$ &  4           & $SO(3,1)$  &  3             & $SO(2,1)$  \\
                  &  2  & $1[12]^2$  &  6           & $SO(5,1)$  &  4             & $SO(3,1)$  \\
                  &  2  & $1[8]^3$   &  8           & $SO(7,1)$  &  5             & $SO(4,1)$  \\ 
                  &  2  & $1[6]^4$   &  10          & $SO(9,1)$  &  6             & $SO(5,1)$  \\
                  &  2  & $1[4]^6$   & 14           & $SO(13,1)$ &  8             & $SO(7,1)$  \\ 
                  &  2  & $1[3]^8$   & 10           & $SO(9,1)$  &  10            & $SO(9,1)$  \\
                  &  2  & $2[2]^{12}$& 26           & $SO(25,1)$ &  14            & $SO(13,1)$  \\ \hline

$\mathbb{Z}_{25}$ &  1  & $20[25]^1$ &  2           & $SO( 1,1)$ &   2            & $SO(1 ,1)$  \\
                  &  1  & $4[5]^5$   &  6           & $SO( 5,1)$ &   6            & $SO( 5,1)$  \\ 
\hline\hline
  \end{tabular}
  \end{center}
  \caption{Some non-prime cyclic groups}
\end{table}


We include two remarks about the data in Table 2: 

\noindent 1) The single-cycle sectors of $\mathbb{Z}_{24}(d=2)$ provide us with $17$ more examples
of $SO(3,1)$-invariant \emph{four-dimensional strings}, whose 
ground-state mass-squared are included in Eq.~(6.4d). See also Sec.~9.

\noindent 2) All the sectors in Table 2 are tachyonic,  except for $\mathbb{Z}_{25}$ -- whose $24$ non-trivial
sectors are entirely non-tachyonic:
\begin{equation}
	\mathbb{Z}_{25}(d=1): \quad 
		P^2(\sigma)_{(0)} =
			\left\{
		\begin{array}{ll}

		\frac{2}{25} & \text{for the 20 single-cycle sectors with } D(\sigma)=2 \\

		0            & \text{for the  4 sectors with } D(\sigma)=6.

		\end{array}
	\right. 
\end{equation}
This is in agreement with our ground-state lemmas in Sec.~5 and the 
single-cycle result in Eq.~(6.3d). See also Sec.~8. The cycle-types for the remaining cyclic groups are given in Table 3. 

\newpage

\begin{table}[h]
\begin{footnotesize} 
  \begin{center}
  \begin{tabular}{c|c|c|c|c||c|c}
\hline\hline
$H(\text{perm})'$	&	$d$	&	cycle-type & $D(\sigma)$	&	symmetry	 &	$D(\sigma)_c$	&	symmetry	\\ \hline

$\mathbb{Z}_{10}$ &	16	&	$[10]^1$ 	 & 18           & $SO(17,1)$ &	17	          &	$SO(16,1)$	\\
                  &	16	&	$[5]^2$ 	 & 18           & $SO(17,1)$ &	18	          &	$SO(17,1)$	\\ 
                  &	16	&	$[2]^5$ 	 & 26           & $SO(25,1)$ &	21	          &	$SO(20,1)$	\\ \hline
               
$\mathbb{Z}_{12}$ & 14  & $[12]^1$  & 16           & $SO(15,1)$ &  15            & $SO(14,1)$  \\
                  & 14  & $[6]^2$   & 18           & $SO(17,1)$ &  16            & $SO(15,1)$  \\
                  & 14  & $[4]^3$   & 20           & $SO(19,1)$ &  17            & $SO(16,1)$  \\ 
                  & 14  & $[3]^4$   & 18           & $SO(17,1)$ &  18            & $SO(17,1)$  \\ 
                  & 14  & $[2]^6$   & 26           & $SO(25,1)$ &  20            & $SO(19,1)$  \\ \hline

$\mathbb{Z}_{14}$ & 12  & $[14]^1$  & 16           & $SO(15,1)$ &  13            & $SO(12,1)$  \\
                  & 12  & $[7]^2$   & 16           & $SO(15,1)$ &  14            & $SO(13,1)$  \\
                  & 12  & $[2]^7$   & 26           & $SO(25,1)$ &  19            & $SO(18,1)$  \\ \hline

$\mathbb{Z}_{15}$ & 11  & $[15]^1$  & 12           & $SO(11,1)$ &  12            & $SO(11,1)$  \\
                  & 11  & $[5]^3$   & 14           & $SO(13,1)$ &  14            & $SO(13,1)$  \\ 
                  & 11  & $[3]^5$   & 16           & $SO(15,1)$ &  16            & $SO(15,1)$  \\ \hline

$\mathbb{Z}_{16}$ &  10  & $[16]^1$ & 12           & $SO(11,1)$  &  11            & $SO(10,1)$  \\
                  &  10  & $[8]^2$  & 14           & $SO(13,1)$  &  12            & $SO(11,1)$  \\
                  &  10  & $[4]^4$  & 18           & $SO(19,1)$  &  14            & $SO(13,1)$  \\ 
                  &  10  & $[2]^8$  & 26           & $SO(25,1)$  &  18            & $SO(17,1)$  \\ \hline

$\mathbb{Z}_{18}$ &  8  & $[18]^1$  &  10          & $SO(9,1)$   &  9             & $SO(8,1)$  \\
                  &  8  & $[9]^2$   &  10          & $SO(9,1)$   &  10            & $SO(9,1)$  \\
                  &  8  & $[6]^3$   &  14          & $SO(13,1)$  &  11            & $SO(10,1)$  \\ 
                  &  8  & $[3]^6$   &  14          & $SO(13,1)$  &  14            & $SO(13,1)$  \\
                  &  8  & $[2]^9$   & 26           & $SO(25,1)$  &  17            & $SO(16,1)$  \\ \hline

$\mathbb{Z}_{20}$ &  6  & $[20]^1$  &  8           & $SO(7,1)$   &  7             & $SO(6,1)$  \\
                  &  6  & $[10]^2$  &  10          & $SO(9,1)$   &  8             & $SO(7,1)$  \\
                  &  6  & $[5]^4$   &  10          & $SO(9,1)$   &  10            & $SO(9,1)$  \\ 
                  &  6  & $[4]^5$   &  16          & $SO(15,1)$  &  11            & $SO(10,1)$  \\
                  &  6  & $[2]^{10}$   & 26        & $SO(25,1)$  &  16            & $SO(15,1)$  \\ \hline

$\mathbb{Z}_{21}$ &  5  & $[21]^1$  &  6           & $SO( 5,1)$ &   6            & $SO(5,1)$  \\
                  &  5  & $[7]^3$   &  8           & $SO( 7,1)$ &   8            & $SO(7,1)$  \\ 
                  &  5  & $[3]^7$   &  12          & $SO(11,1)$ &   12           & $SO(11,1)$  \\ 
\hline\hline
  \end{tabular}
  \end{center}
  \end{footnotesize}
  \caption{The rest of the cyclic groups}
\end{table}

\underline{Systematics of the permutation groups}

We now consider systematic statements  for the target space-time 
dimensionalities $\{ D(\sigma)\}$ associated to the general permutation group $H(\text{perm})'_{26-d}$.

Surveying the data in the tables above for all the cyclic groups $H(\text{perm})'_{26-d} = \mathbb{Z}_{26-d}$, one  
checks explicitly for these groups that 
\begin{subequations}
\begin{equation}
 D(\sigma) = d + N_O(\sigma)' + 2 N_E(\sigma)',\quad 1\leq d\leq 25
\end{equation}
\begin{equation}
 2\leq (D(\sigma) = \text{even}) \leq 26, \quad 1\leq d\leq 26.
\end{equation}
\end{subequations}
The trivial elements ($D(\sigma)=26$) of each cyclic group and the case of no permutation group $D(\sigma) = d = 26$ is included in the 
double inequality (7.6b).

We remind [8] that the formula for $D(\sigma)$ in Eq.~(7.6a) holds for all permutation groups 
$H(\text{perm})'_{26-d}$, and we know from the counting in 
Eq.~(4.9) that $D(\sigma) \leq 26$ for the general case. In fact, it is not difficult to see that
the full double inequality (7.6b), including $D(\sigma)=$ even, holds as 
well for all  $H(\text{perm})'_{26-d}$.

We sketch this first for the symmetric groups $H(\text{perm})'_{26-d} = S_{26-d}$, where 
the sectors correspond to the ordered partitions of $(26-d)$:
\begin{equation}
  S_{26-d}: \quad \sum_L F_L(\sigma) = 26-d, \quad 1\leq F_{L+1}(\sigma) \leq F_L(\sigma), \quad 1\leq d\leq 25.
\end{equation}
As detailed above for the cyclic groups, this description allows the 
inspection of the group elements of the 24 relevant non-trivial 
symmetric groups $S_{2}(d=24)...S_{25}(d=1)$, leading to the following observations:

\noindent a) For each symmetric group, the minimum space-time dimension is achieved for the 
single-cycle sectors (see Sec.~6)  with
\begin{equation}
 D(\sigma) = 
\left\{    
		\begin{array}{ll}
		d+2 & \text{for $d$ even}  \\
		d+1 & \text{for $d$ odd}. 
		\end{array}
\right.
\end{equation}
This provides the lower bound $D(\sigma) \geq 2$ in Eq.~(7.6b) for the 
symmetric groups, and the equality is
achieved among these groups only by the single-cycle elements of $S_{25}(d=1)$.

\noindent b) In each sector of each symmetric group, the number of cycles 
of odd length $F_L(\sigma)$ satisfies
\begin{equation}
  N_{0}(\sigma)' = 
\left\{    
		\begin{array}{l}
		\text{even for } d \text{ even}  \\
		\text{odd for } d \text{ odd} 
		\end{array}
\right.
\end{equation}
and therefore 
\begin{subequations}
\begin{equation}
	d+ N_O(\sigma)' = \text{even for all $d$}
\end{equation}
\begin{equation}
	D(\sigma) = \text{even for all $d$}.
\end{equation}
\end{subequations}

\noindent c) The maximum space-time dimension for each symmetric group is $D(\sigma) = 26$, e.g.
\begin{equation}
	D(\sigma) = 26: \quad [2]^{\frac{26-d}{2}} \text{ for $d$ even},\,\, [2]^{\frac{25-d}{2}}[1]^1 \text{ for $d$ odd}
\end{equation}
which provides the last part of the double inequality (7.6b) for the 
symmetric groups.

Finally, since all permutation groups are subgroups of the symmetric groups
\begin{equation}
	H(\text{perm})'_{26-d} \subset S_{26-d}
\end{equation}
we conclude that the double inequality (7.6b) holds for all 
$H(\text{perm})'_{26-d}$. 

The double inequality (7.6b) is one of the central results of this paper, holding  across all 
open-string sectors in the large example (1.5) of orientation-orbifold 
string systems. 
We shall return to some physical consequences of this result below. 

A final remark is essential here. As seen explicitly in the tables above, the 
dimension $D(\sigma)$ of these open-string Lorentzian space-times generally vary from 
sector to sector even in the same orbifold. It follows that the 
as-yet-unconstructed \emph{twist-fields} (intertwiners) of these orbifolds will 
characteristically induce \emph{target space-time transitions} 
$\Delta(D(\sigma)) \neq 0$.

\section{The Non-Tachyonic Strings}

\noindent We have noted in Sec.~5 and Eq.~(7.5) that the only non-tachyonic 
open-string sectors in $\{H(\text{perm})'_{26-d} = \mathbb{Z}_{26-d}\}$ are all $24$ non-trivial
sectors  $\sigma = 1,\dots, 24$ of $\mathbb{Z}_{25}$ ($d=1$) (see also 
Table 2). For future reference, we write out these sectors here in further detail.

\underline{20 sectors of type $[25]^1$ with $SO(1,1)$ symmetry}

\begin{subequations}
\begin{multline}
  L(M) = \frac{26}{25} \delta_{M,0} - \frac{1}{2} \eta_{(2)}^{ab} \sum_{Q\in\mathbb{Z}} 
    \normalorder J_a(Q) J_b(M-Q)\normalorder_M + \\
    + \frac{1}{50} \sum_{\hat L=1}^{24} \sum_{Q\in\mathbb{Z}}
    \normalorder J_{\hat L 0}(Q+\frac{2\hat L}{25}) J_{-\hat L, 0}(M-Q-\frac{2\hat L}{25}) \normalorder_M
\end{multline}
\begin{equation}
	P^2(\sigma) = \eta_{(2)}^{ab} J_a(0) J_b(0)
\end{equation}
\begin{equation}
	D(\sigma) = 2, \quad P^2(\sigma)_{(0)} = \frac{2}{25},\quad 1\leq 
	(\forall \sigma \text{ not a multiple of $5$}) \leq 24
\end{equation}
\begin{equation}
	\Delta(P^2(\sigma)) = 
\left\{    
		\begin{array}{ll}
		2 |M| & \text{for } J_a(M<0),\,\, a = 0,1  \\
		2 |M + \frac{2 \hat L}{25}| & \text{for } J_{\hat L 0}((M+\frac{2\hat L}{25})<0),\,\, \hat L = 1,\dots,24.  
		\end{array}
\right.
\end{equation}
\end{subequations}
In this result, the time dimension comes from $d=1$ while the single spatial 
dimension comes from the old $\hat{L}=0$ current.

\underline{4 sectors of type $[5]^5$ with $SO(5,1)$ symmetry}

\begin{subequations}
\begin{multline}
	L(M) = \delta_{M,0} - \frac{1}{2} \eta^{ab}_{(6)} \sum_{Q\in\mathbb{Z}} 
	\normalorder J_a(Q) J_b(M-Q)\normalorder_M +\\
		+ \frac{1}{10} \sum_{L=0}^4 \sum_{\hat L=1}^4 \sum_{Q\in\mathbb{Z}} 
			\normalorder J_{\hat L L}(Q+\frac{2\hat L}{5}) J_{-\hat L,L}(M-Q-\frac{2\hat L}{5}) \normalorder_M
\end{multline}
\begin{equation}
	P^2(\sigma) = \eta^{ab}_{(6)} J_a(0) J_b(0)
\end{equation}
\begin{equation}
	D(\sigma) = 6,\quad P^2(\sigma)_{(0)} = 0 , \quad \sigma = 5,10, 15,20
\end{equation}
\begin{equation}
	\Delta(P^2(\sigma)) = 
\left\{    
		\begin{array}{ll}
		2 |M| & \text{for } J_a(M<0),\,\, a = 0,1, \dots, 5  \\
		2 |M + \frac{2 \hat L}{5}| & \text{for } J_{\hat L L}((M+\frac{2\hat L}{5})<0),\,\,  L = 0,1,\dots,4,\,\, \hat L = 1,\dots, 4.
		\end{array}
\right.
\end{equation}
\end{subequations}
In these sectors the time dimension comes again from $d=1$, while the 
five spatial dimensions come from the five old $\hat{L}=0$ currents.
There remains for $\mathbb{Z}_{25}$ the trivial open-string sector $\sigma = 0$ with space-time
dimension 26, which is spectrally- equivalent to the ordinary critical open string.

\section{The Four-Dimensional Strings}

\noindent In this section we collect all the $SO(3,1)$-invariant four-dimensional 
open strings associated 
with $\{ H(\text{perm})'_{26-d} = \mathbb{Z}_{26-d} \}$.

\underline{All non-trivial sectors of $H(\text{perm})'_{23} = \mathbb{Z}_{23}$}

Since 23 is prime, $\mathbb{Z}_{23}$ has 22 non-trivial sectors of single-cycle
type $[23]^1$ --- all of which are four dimensional strings (see Table 1).
In these cases, we include further details for the currents:
\begin{subequations}
\begin{multline}
	L(M) = \frac{22}{23} \delta_{M,0} - \frac{1}{2} \eta^{ab}_{(4)}
		\sum_{Q\in\mathbb{Z}} \normalorder J_a(Q) J_b(M-Q) \normalorder_M +\\
		+ \frac{1}{46} \sum_{\hat L=1}^{22} \sum_{Q\in\mathbb{Z}} 
			\normalorder \hat J_{\hat L 0}(Q+\frac{2\hat L}{23}) J_{-\hat L,0} (M-Q-\frac{2\hat L}{23}) \normalorder_M
\end{multline}
\begin{equation}
	[ L(M),J_a(N)] = -N J_a(M+N), \quad a=0,1,2,3
\end{equation}
\begin{equation}
	[J_a(M), J_b(M) ] = \eta_{(4)}^{ab} N \delta_{M+N,0}
\end{equation}
\begin{equation}
	[L(M), J_{\hat L0}(N+\frac{2\hat L}{23}) ] = - (N+\frac{2\hat L}{23}) J_{\hat L0}(M+N+\frac{2\hat L}{23}), \quad \hat L=1,\dots,22
\end{equation}
\begin{equation}
	[J_{\hat J0}(M+\frac{2\hat J}{23}), J_{\hat L0}(N+\frac{2\hat L}{23}) ] = (23 M+2\hat L) \delta_{\hat J+\hat L,0\text{ mod } 23} \delta_{M+N+2(\frac{\hat J + \hat L}{23}),0}
\end{equation}
\begin{equation}
	d=3, \quad D(\sigma)=4, \quad P^2(\sigma)_{(0)} = -\frac{2}{23}, \quad \sigma =1, \dots, 22
\end{equation}
\begin{equation}
	P^2(\sigma) =\eta^{ab}_{(4)} J_a(0)J_b(0),\quad \Delta(P^2(\sigma)) =
	\left\{    
		\begin{array}{ll}
		2 |M| & \text{for } J_a(M<0),\, a = 0,1,2,3  \\
		2 |M + \frac{2 \hat L}{23}| & \text{for } J_{\hat L 0}((M+\frac{2\hat L}{23})<0),\, \hat L = 1,\dots,22.
		\end{array}
	\right.
\end{equation}
\end{subequations}
We remind that the integer-moded currents $\{ J_a(M), a=0,1,2,3 \}$ 
transform as Lorentz four-vectors under $SO(3,1)$, while
the extra 22 fractional-moded currents $\{ J_{\hat L0}(M + \frac{2\hat L}{23}), \hat L=1, \dots, 22 \}$ are
scalars under the four-dimensional Lorentz group.  The remaining 
(trivial) open-string sector $\sigma=0$ of $\mathbb{Z}_{23}$ has, as usual, $D(\sigma)=26$ and is 
spectrally-equivalent to the untwisted critical open string.

\underline{The 17 single-cycle sectors of $H(\text{perm})'_{24} = \mathbb{Z}_{24}$}

The 17 single-cycle sectors $[24]^1$ of $\mathbb{Z}_{24}$ are also
four-dimensional strings (see Table 2):
\begin{subequations}
\begin{multline}
	L(M) = \frac{143}{144} \delta_{M,0} -
		\frac{1}{2} \eta_{(4)}^{ab} \sum_{Q\in\mathbb{Z}}
			\normalorder J_a(Q)J_b(M-Q) \normalorder_M  +\\
			+\frac{1}{34} \sum_{\hat L\neq 0,12}^{23} \sum_{Q\in\mathbb{Z}}
			\normalorder J_{\hat L0}(Q+\frac{\hat L}{12}) J_{-\hat L,0}(M-Q-\frac{\hat L}{12})\normalorder_M
\end{multline}
\begin{equation}
	[L(M), J_a(N)] = -N J_a(M+N), \quad a=0,1,2,3
\end{equation}
\begin{equation}
	[J_a(M), J_b(N)]=\eta_{(4)}^{ab} N \delta_{M+N,0}
\end{equation}
\begin{equation}
	[L(M), J_{\hat L0}(N+\frac{\hat L}{12}) ] = -(N+\frac{\hat L}{12}) J_{\hat L0}(M+N+\frac{\hat L}{12}), \quad \hat L=1\dots 11 \text{ and } 13\dots 23
\end{equation}
\begin{equation}
	[J_{\hat J0}(M+\frac{\hat J}{12}), J_{\hat L0}(N+\frac{\hat L}{12}) ] = 2(12 M + 2\hat L) \delta_{\hat J+\hat L, 0\text{ mod } 24} \delta_{M+N+\frac{\hat J+\hat L}{12},0}
\end{equation}
\begin{equation}
	d=2, \quad D(\sigma)=4, \quad P^2(\sigma)_{(0)} = -\frac{1}{72}, \quad \sigma \text{ not a divisor of } 24
\end{equation}
\begin{equation}
	P^2(\sigma) = \eta^{ab}_{(4)} J_a(0) J_b(0), \quad \Delta(P^2(\sigma)) = 
	\left\{    
		\begin{array}{ll}
		2 |M| & \text{for } J_a(M<0),\, a = 0,1,2,3  \\
		2 |M + \frac{\hat L}{12}| & \text{for } J_{\hat L 0}((M+\frac{\hat L}{12})<0), \,\hat L \neq 0,12.
		\end{array}
	\right.
\end{equation}
\end{subequations}
Note that here, as in the sectors of $\mathbb{Z}_{23}$ above, there are 
exactly 22 extra fractional-moded Lorentz scalar fields, i.e. a total of $c(\sigma) = 4+22= 26$ effective
degrees of freedom in sector $\sigma$. 
As seen in Table 2, the multi-cycle sectors of $\mathbb{Z}_{24}$ have 
higher space-time dimensions $D(\sigma) = 6,8,10,14$ and $26$ with 
correspondingly larger Lorentz symmetries $SO(D(\sigma)-1,1)$.

\section{Physical States: \\\quad The First Four Levels of $H(\text{perm})'_3=\mathbb{Z}_3$}

\noindent In this and the following three sections, we present some introductory remarks 
on the \emph{physical states} of the open-string sectors in the 
orientation-orbifold string systems.

We begin here with the low-lying states of the two non-trivial open-string
sectors $\sigma = 1,2$ of the simplest case $H(\text{perm})'_3=\mathbb{Z}_3$ (see Table 1). These 
strings are described by the following dynamics in both sectors:
\begin{subequations}
\begin{equation}
	(L(M\geq 0) - \delta_{M,0}) | \chi (\sigma)\rangle = 0, \quad \sigma = 1,2
\end{equation}
\begin{multline}
	L(M) = \frac{1}{9} \delta_{M,0} - \frac{1}{2} \eta^{ab}_{(24)} \sum_{Q\in\mathbb{Z}}
		\normalorder J_a(Q) J_b(M-Q) \normalorder_M +\\
		+ \frac{1}{6} \sum_{\hat L=1}^2 \sum_{Q\in\mathbb{Z}} 
			\normalorder J_{\hat L0}(Q+\frac{2\hat L}{3}) J_{-\hat L, 0}(M-Q-\frac{2\hat L}{3})\normalorder_M
\end{multline}
\begin{equation}
	[L(M),L(N)] = (M-N) L(M+N) + \frac{26}{12} M(M^2-1) \delta_{M+N,0}
\end{equation}
\begin{equation}
	[L(M),J_a(N)] = -N J_a(M+N),\quad a=0,1,\dots, 23
\end{equation}
\begin{equation}
	[J_a(M),J_b(N)] = \eta^{(24)}_{ab} N\delta_{M+N,0}
\end{equation}
\begin{equation}
	[L(M),J_{\hat L0}(N+\frac{2\hat L}{3})] = -(N+\frac{2\hat L}{3}) J_{\hat L0}(M+N+\frac{2\hat L}{3}), \quad \hat L=1,2
\end{equation}
\begin{equation}
	[J_{\hat J0}(M+\frac{2\hat J}{3}), J_{\hat L0}(N+\frac{2\hat L}{3})]
		= (3M+2\hat L)\delta_{\hat J+\hat L, 0\text{ mod }3}\delta_{M+N+2\frac{\hat J+\hat L}{3},0}
\end{equation}
\begin{equation}
	d=23,\quad D(\sigma)=24,\quad P^2(\sigma) = \eta^{ab}_{(24)} J_a(0) J_b(0), \quad P^2(\sigma)_{(0)} = -\frac{16}{9}
\end{equation}
\begin{equation}
	\Delta(P^2(\sigma)) = 2 | \text{mode \#}| \,\,\,\,\text{ for each } J (\text{(mode \#)}<0)
\end{equation}
\begin{equation}
	J_a(M)^\dagger = J_a(-M), \quad J_{\hat L0}(M+ \frac{2\hat L}{3})^\dagger = J_{-\hat L,0}(-M - \frac{2\hat L}{3})
\end{equation}
\begin{equation}
	J_{\hat L\pm 3, 0}(M+\frac{2(\hat L\pm 3)}{3}) = J_{\hat L0}(M\pm 2 + \frac{2\hat L}{3}).
\end{equation}
\end{subequations}
Each of these sectors is an $SO(23,1)$-invariant 24-dimensional string, 
where the integer-moded currents $\{ J_a, a=0,\dots, 23 \}$ transform
as Lorentz vectors  and the extra two fractional-moded currents $\{ J_{\hat L0},\hat L = 1,2 \}$ are Lorentz scalars.

The physical states of each sector $\sigma$ are the solutions of
the physical-state condition in Eq.~(10.1a). The space of physical states
can be analyzed as usual at fixed values of the \emph{level}
\begin{equation}
	\text{level} \equiv \text{total mode \#} \equiv \sum |\text{mode \# of the currents}| 
\end{equation}
a fact which holds as well in any sector of all the orbifold-string 
theories of permutation-type.

Among the Lorentz scalars, one finds that there is exactly one negative,fractional
mode of each type
\[
	J_{10}(-\frac{1}{3}), J_{20}(-\frac{2}{3}), J_{10}(-\frac{4}{3}), J_{20}(-\frac{5}{3}), J_{10}(-\frac{7}{3}), J_{20}(-\frac{8}{3}), \dots
\]
and, owing to the periodicity conditions (10.1k), modes with $\hat L = 
-1, -2$ are equivalent to entries in this list.

After some algebra, we obtain the following list of all physical states 
for the first four levels of these sectors:

\emph{level 0}:
\begin{equation}
	|P^2(\sigma) \!=\! P^2(\sigma)_{(0)}\! =\! -\frac{16}{9} \rangle
\end{equation}

\emph{level 1/3}:
\begin{subequations}
\begin{equation}
	J_{10}(-1/3) | P^2(\sigma)\! =\! -\frac{13}{9} \rangle
\end{equation}
\begin{equation}
	|| J_{10}(-1/3) | P^2(\sigma)\! =\! -\frac{13}{9}\rangle ||^2 = ||\,\, |P^2(\sigma)\! = \!\frac{13}{9} \rangle ||^2 
\end{equation}
\end{subequations}

\emph{level 2/3}:
\begin{subequations}
\begin{equation}
	J_{20}(-2/3) | P^2(\sigma)\! =\! -\frac{4}{9} \rangle
\end{equation}
\begin{equation}
	|| J_{20}(-2/3) | P^2(\sigma) \!= \!-\frac{4}{9}\rangle ||^2 = ||\,\, |P^2(\sigma)\! = \!-\frac{4}{9} \rangle ||^2 
\end{equation}
\begin{equation}
	J_{10}(-1/3)J_{10}(-1/3) | P^2(\sigma) \!= \!-\frac{4}{9}\rangle
\end{equation} 
\begin{equation}
	|| J_{10}(-1/3)J_{10}(-1/3) | P^2(\sigma)\! =\! -\frac{4}{9}\rangle ||^2 = 2 ||\, |P^2(\sigma)\! =\! -\frac{4}{9} \rangle ||^2 
\end{equation}
\end{subequations}

\emph{level 1}:
\begin{subequations}
\begin{equation}
	\epsilon(\sigma)\! \cdot\! J(-1) | P^2(\sigma)\! = \!\frac{2}{9} \rangle, \quad \epsilon(\sigma)\!\cdot\! P(\sigma) = 0, \quad \epsilon^2(\sigma)=-1
\end{equation}
\begin{equation}
	|| \epsilon(\sigma)\! \cdot\! J(-1) | P^2(\sigma) \!= \!\frac{2}{9}\rangle ||^2 = ||\,\, |P^2(\sigma) \!= \!\frac{2}{9} \rangle ||^2 
\end{equation}
\begin{equation}
	L(-1) | P^2(\sigma)\! = \!\frac{2}{9} \rangle  
	= (-P(\sigma)\! \cdot\! J(-1) + \frac{1}{3} J_{10}(-\frac{1}{3}) J_{20}(-\frac{2}{3})) |P^2(\sigma)\! =\!\frac{2}{9}\rangle
\end{equation}
\begin{equation}
	|| L(-1) |P^2(\sigma)\! = \!\frac{2}{9} \rangle ||^2 = 0
\end{equation}
\end{subequations}
In these results, we have introduced an alternate notation for the momentum eigenstates
\begin{equation}
	|P^2(\sigma)\rangle \equiv |0, J(0)_\sigma \rangle
\end{equation}
to conform with standard usage in string theory. The inner products above 
are defined with the 24-dimensional Lorentz metric
\begin{equation}
	A(\sigma)\! \cdot\! B(\sigma) = \eta^{ab}_{(24)} A_a(\sigma) B_b(\sigma) = A_0(\sigma) B_0(\sigma) - \vec{A}(\sigma)\!\cdot\!\vec{B}(\sigma)
\end{equation}
where the momenta $\{P_a(\sigma) = J_a(0)\}$ are the zero modes of the 
sector. The last result in Eq.~(10.6d) follows either from the Virasoro 
algebra or the current algebra.

The $SO(23,1)$ spin of these physical states is easily read as follows
\begin{subequations}
\begin{align}
\text{level }  0   & : \text{tachyonic ground state at } J = 0 \\
\text{level }  1/3 & : \text{higher tachyon at } J = 0 \\
\text{level }  2/3 & : \text{two still higher tachyons at } J = 0 \\
\text{level }  1   & : \text{non-tachyonic states at } J = 1, 0
\end{align}
\end{subequations}
where $J=1$ and 0 denote Lorentz vectors and scalars respectively, as is 
conventional in string theory. More generally for space-time dimension $D(\sigma)$ with $SO(D(\sigma)-1,1)$ symmetry,
the ground-state lemmas of Sec.~5 and the presence of the $(26-D(\sigma))$ Lorentz scalars means that the 
physical spectra of these bosonic prototypes often begins with a short 
set of tachyons (see however Sec.~8).

The oscillator-free momentum eigenstates $\{ | P^2(\sigma) \rangle \}$ 
have positive norm as usual, so  there are no negative-norm states in the 
first four levels of $H(\text{perm})'_3 =\mathbb{Z}_3$.

Moreover, the only negative-norm basis states in each sector are
(as usual) those with an odd number of time-like modes $\{ J_0(M<0) \}$. 
In particular, there are no negative norms in the positive-definite 
Hilbert space spanned by the Lorentz-scalar fractionally-moded currents $\{ J_{\hat L0}((M+\frac{2\hat L}{3})<0), \hat L=1,2 \}$.  
These facts are so far consistent with the no-ghost conjecture (see 
Refs.[1,4,5,7,8] and Sec.~12) 
for each Lorentzian sector of all the orbifold-string theories of permutation-type.

On the other hand, we have already found in Eqs.~(10.6c,d) a zero-norm physical state at 
level 1, whose decoupling will depend on the gauges [4,5] of the interacting 
theory. We shall return to this part of the no-ghost conjecture 
in Sec.~12.

\section{The $\hat c(\sigma)=52$ Description of the Physical States}

\noindent In the previous section, we solved the reduced  $c(\sigma)=26$ 
formulation for the low-lying physical states (10.3)-(10.6) of the sectors 
$\sigma=1,2$ associated to the non-trivial elements of $H(\text{perm})'_{3}$.
 Solving for the physical states  was straightforward in the reduced 
 description (4.5c), 
at least in part because the enhanced Lorentz transformation properties 
of the currents and states are transparent in this formulation. It is 
instructive however to reconsider these same physical states as well in the original 
description of these sectors at $\hat c(\sigma)=52$.

In the unreduced $\hat c(\sigma)=52$ description of the same sectors, we 
remind that the physical states are
the solutions to the extended Virasoro conditions (2.3a), 
using the orbifold Virasoro generators
\begin{multline}
	\hat L_{\hat j}(m + \frac{\hat j}{2}) =\, \frac{125}{72} \delta_{m+\frac{\hat j}{2},0}
		- \frac{1}{4} \eta^{ab}_{(23)}  \sum_{\hat \ell=0}^1 \sum_{p\in\mathbb{Z}} 
			\normalorder \hat J_{\epsilon a\hat \ell} (p + \frac{\epsilon+\hat \ell}{2}) \hat J_{-\epsilon,b,\hat j-\hat\ell} (m - p + \frac{\hat j -\hat\ell -\epsilon}{2}) \normalorder_M +\\
        + \frac{1}{12} \sum_{\hat\ell=0}^1\sum_{\hat L=0}^2 \sum_{p\in\mathbb{Z}} 
			\normalorder \hat J_{\hat L 0\hat\ell}(p + \frac{\hat L}{3} + \frac{\hat \ell}{2}) \hat J_{-\hat L, 0, \hat j-\hat\ell} (m - p - \frac{\hat L}{3} + \frac{\hat j -\hat \ell-\epsilon}{2}) \normalorder_M
\end{multline}
and the hatted current algebras (2.5) with $F_L(\sigma) =3$, $L=0$. In 
this unreduced form,  
we cannot see clearly either the ``preliminary'' Lorentz symmetry $SO(22,1)$ from 
$d=23$ (masked in the second term), or the last spatial dimension (masked in the last term) of the enhanced Lorentz symmetry 
$SO(23,1)$. 
 
Alternately however, the $\hat c(\sigma)=52$ description of the physical states in 
question can  be straightforwardly obtained from the $c(\sigma) = 26$ description of the
states (10.3)-(10.6) by using the inverse of the map (3.1) 
and the relabeling definitions (4.3) and (4.4).

This gives the first four levels  of $H(\text{perm})'_3=\mathbb{Z}_3$ in the $\hat c(\sigma) = 52$ description:
\begin{subequations}
	\begin{equation}
		| P^2(\sigma)\! = \!-\frac{16}{9} \rangle = | \hat P^2(\sigma)\! =\! -\frac{16}{9}\rangle
	\end{equation}
	\begin{equation}
		J_{10}(-\frac{1}{3}) | P^2(\sigma)\! =\! -\frac{13}{9}\rangle  = \hat J_{101} (-\frac{1}{6}) | \hat P^2(\sigma)\! = \!-\frac{13}{9}\rangle
	\end{equation}
	\begin{equation}
		J_{20}(-\frac{2}{3}) | P^2(\sigma)\! =\! -\frac{4}{9}\rangle  = \hat J_{200} (-\frac{1}{3}) | \hat P^2(\sigma) \!=\! -\frac{4}{9}\rangle
	\end{equation}
	\begin{equation}
		\epsilon(\sigma) \!\cdot \!J(-1) |P^2(\sigma)\! = \!\frac{2}{9}\rangle = 
	(\sum^{22}_{a=0} \epsilon^a(\sigma) \hat J_{\epsilon a,1-\epsilon} (-\frac{1}{2}) 
	+ \frac{1}{\sqrt{3}} \epsilon^{23}(\sigma) \hat J_{001}(-\frac{1}{2})) |\hat P^2(\sigma)\! = \!\frac{2}{9}\rangle
	\end{equation}
	\begin{equation}
		0 = \epsilon(\sigma) \!\cdot\! P(\sigma) = \sum_{a=0}^{22} \epsilon^a(\sigma) \hat J_{\epsilon a \epsilon}(0) + \frac{1}{\sqrt{3}} \epsilon^{23}(\sigma) \hat J_{000}(0), \quad \epsilon^2(\sigma)=-1
	\end{equation}
	\begin{equation}
		L(-1) |P^2(\sigma)\!=\!\frac{2}{9}\rangle = 2\hat L_1(-\frac{1}{2}) |\hat P^2(\sigma) \!= \!\frac{2}{9}\rangle .
	\end{equation}
\end{subequations}
Note that the momenta of the states are invariant under the map, but the total
mode number is multiplied  by $1/2$, so that the first four levels in the 
$\hat c(\sigma)=52$ description are $0$, $1/6$, $1/3$ and $1/2$.

We add three observations at this point:

\noindent 1) As discussed in Refs.~[7,8] and Sec.~4, the target space-time structure 
of each j-cycle of the orbifolds of permutation-type
is preserved by the map, including  in particular the space-time 
dimensions $D(\sigma)$ and the enhanced Lorentz symmetry $SO(D(\sigma)-1,1)$
-- though the space-time structure is masked in the $\hat c(\sigma)=52$ formulation.
In the present case for example, the structures
\begin{subequations}
\begin{equation}
	J_a(-1) = \left\{
		\begin{array}{ll}

		\hat J_{\epsilon a,1-\epsilon}(-\frac{1}{2}) & \text{for } a = 0,1,\dots, 22 \\

		\frac{1}{\sqrt{3}} \hat J_{001}(-\frac{1}{2})      & \text{for } a = 23

		\end{array}
	\right.
\end{equation}
\begin{equation}
	J_a( 0) = \left\{
		\begin{array}{ll}

		\hat J_{\epsilon a\epsilon}(0)    & \text{for } a = 0,1,\dots, 22 \\

		\frac{1}{\sqrt{3}} \hat J_{000}(0) & \text{for } a = 23

		\end{array}
	\right.
\end{equation}
\end{subequations}
are Lorentz 24-vectors while the modes $\hat J_{101}(-\frac{1}{6})$ and 
$\hat J_{200}(-\frac{1}{3})$ are still scalars under $SO(23,1)$.
Although we will not pursue this here, the enhanced space-time symmetry $SO(23,1)$ of the
generators (11.1) can be made manifest by field-redefinitions closely 
related to the examples in Eq.~(11.3).

\noindent 2) As noted more generally in Ref.~[8] and Sec.~3, the norms 
and inner products of all states in the $\hat c(\sigma)=52$ description
are the same as those computed in the reduced description at $c(\sigma)=26$.

\noindent 3) In particular, the zero-norm state in (Eqs.~(10.6c,d) and) Eq.~(11.2f) is still 
associated with the orbifold Virasoro generators. This fact will play a 
role in the discussion of the following section.

\section{The No-Ghost Conjecture}

\noindent We have conjectured [1,4,5,8] a no-ghost theorem for all the 
Lorentzian [8] orbifold-string
theories of permutation-type, including in particular all the sectors of 
the large example (1.5) of orientation-orbifold
string systems. This conjecture is based on two observations:

\noindent 1) The untwisted sectors $U(1)^{26K}$ or $(U(1)^{26}_L \times U(1)^{26}_R)$ of the 
orbifolds of permutation-type are free of negative-norm and zero-norm states .

\noindent 2) The twisted sectors of the orbifold-string theories of 
permutation-type are constructed from these untwisted sectors, following 
known principles of orbifold theory [9-23].

\noindent On this basis, one might expect that an orbifoldization in light-cone gauge could
establish the no-ghost theorem for these theories -- but such an approach 
has not yet been attempted, and could
shroud a number of issues, such as the new world-sheet geometries [1]
and the enhanced target space-time symmetries [8] of the new string theories.

We outline here several parts of the conjecture in our covariant construction
of the orientation-orbifold string systems:

\noindent a) \emph{No negative-norm physical states}\\
We expect that the extended physical-state conditions (2.3a) at $\hat c(\sigma) = 52$
\begin{subequations}
\begin{equation}
	(\hat L_{\hat j}((m+\frac{\hat j}{2})\geq 0) - \hat a_2 \delta_{m+\frac{\hat j}{2},0} ) | \chi(\sigma) \rangle = 0
\end{equation}
\begin{equation}
	\bar{\hat j} = 0,1,\quad \hat a_2 = \frac{17}{8}
\end{equation}
\end{subequations}
or the equivalent, reduced physical-state condition (3.2a)  at $c(\sigma)=26$
\begin{equation}
	(L(M\geq 0) - \delta_{M,0}) | \chi(\sigma)\rangle = 0 , \quad M = 2m+\bar{\hat j}
\end{equation}
will eliminate all negative-norm states from the physical spectrum of each sector $\sigma$.

\noindent b) \emph{Null physical states}\\
We expect that, as in ordinary open-string theory, \emph{all} null 
physical states will be linear combinations of mass-shell states of the form
\begin{subequations}
\begin{equation}
	2\hat L_{- \hat j}(-m-\frac{\hat j}{2}) | \hat L_0(0) = \hat a_2 - m - 
	\frac{\hat j}{2}\rangle = L(-M) | L(0)=1-M\rangle
\end{equation}
\begin{equation}
	M = 2m + \bar{\hat j} = 1,2, \dots .
\end{equation}
\end{subequations}
We have encountered a null physical state of this form in Sections 10 
and 11.
Note that the reduced forms in Eqs.~(12.2) and (12.3a) are the \emph{same} as ordinary
open-string theory. Moreover, as in ordinary string theory, all physical 
states of the form (12.3) are  null physical states.

\noindent c) \emph{Conjectured form of the gauges}\\
We conjecture that \emph{all} null physical states will decouple by a \emph{single 
family} of gauges whose form is expected to be
\begin{subequations}
\begin{equation}
	\hat W_{\hat j} (( m + \frac{\hat j}{2})>0) \equiv \hat \beta (m+ \frac{\hat j}{2}) \hat L_{\hat j}(m+\frac{\hat j}{2}) - (\hat L_0(0) - \hat a_2 + m + \frac{\hat j}{2})
\end{equation}
\begin{equation}
	= \frac{1}{2} W(M>0)
\end{equation}
\begin{equation}
	W(M) = \beta(M) L(M) - (L(0) - 1+M)
\end{equation}
\end{subequations}
where the $\beta$'s are model-dependent non-zero constants satisfying 
\begin{equation}
	\hat\beta_{\hat j}(m+\frac{\hat j}{2})^* =  \beta_{-\hat j}(-m -\frac{\hat j}{2}),
	\quad \beta(M) \equiv \hat \beta_{\hat j}(m + \frac{\hat j}{2}),\quad \beta(M)^* = \beta(-M).
\end{equation}
It is sraightforward to see that these gauges satisfy the following relations:
\begin{subequations}
\begin{equation}
	\hat W_{\hat j} ((m+\frac{\hat j}{2})>0)^\dagger | \hat L_0(0) = \hat a_2 -m -\frac{\hat j}{2}\rangle
\end{equation}
\begin{equation}
	\,\,\,\,\,\quad\,\,\,\,\, = \hat \beta_{-\hat j}(-m-\frac{\hat j}{2}) \hat L_{\hat j} (-m-\frac{\hat j}{2}) | \hat L_0(0)= \hat a_2-m-\frac{\hat j}{2} \rangle
\end{equation}
\begin{equation}
	= \frac{1}{2} W(M>0)^\dagger |L(0) = 1-M\rangle
\end{equation}
\begin{equation}
	= \frac{1}{2}\beta(M) L(-M) |L(0) = 1-M\rangle.
\end{equation}
\end{subequations}
In fact such gauges $\{ \hat W_{\hat j} \}$, with non-trivial 
$\hat\beta$, have been found explicitly [4] in a simple case of the
orientation-orbifold string theories (see also Ref.~[5]), and the next step would be
to check the ``activity'' of these gauges vis-a-vis the more general 
twisted vertex operators outlined for the $\hat{c}=52$ formulation in the Appendix of 
Ref.~[4].

It is clear that parts (a) and (b) of the conjecture are true for 
our examples above in $H(\text{perm})'_3=\mathbb{Z}_3$ .

Moreover, it is easily checked that parts (a) and (b) are true for all sectors $\sigma$
thru level one for any choice $H(\text{perm})'_{26-d}$ in the 
orientation-orbifold string systems: For all levels less than one, the 
physical states are formed entirely from the extra $(26-D(\sigma))$ Lorentz-scalar 
fractional-moded currents, with positive-definite norms, while the only 
physical states at level one are
\begin{subequations}
\begin{equation}
	\epsilon(\sigma)\!\cdot\! J(-1) |P^2(\sigma) \!=\! P^2(\sigma)_{(0)} \!+ 2\rangle, \quad \epsilon(\sigma)\!\cdot\! P(\sigma) = 0, \quad \epsilon^2(\sigma) = -1
\end{equation}
\begin{equation}
	L(-1) |P^2(\sigma) \!=\! P^2(\sigma)_{(0)} \!+ 2\rangle .
\end{equation}
\end{subequations}
The norms of the level-one physical states are respectively positive and zero
\begin{subequations}
\begin{equation}
	|| \epsilon(\sigma)\!\cdot \!J(-1) |P^2(\sigma) \!=\! P^2(\sigma)_{(0)} \!+ 2\rangle ||^2 = ||\,\, | P^2(\sigma)\! = \!P^2(\sigma)_{(0)} \!+ 2 \rangle ||^2
\end{equation}
\begin{equation}
	|| L(-1) |P^2(\sigma)\! =\! P^2(\sigma)_{(0)} \!+ 2 \rangle ||^2 = \langle P^2(\sigma) | 2L(0) | P^2(\sigma) \rangle
\end{equation}
\begin{equation}
	= (-P^2(\sigma) + 2\hat\delta_0(\sigma)) ||\,\, |P^2(\sigma) \!=\! P^2(\sigma)_{(0)} \!+ 2 \rangle ||^2 = 0
\end{equation}
\end{subequations}
for all open-string sectors $\sigma$ of all our theories, generalizing 
our $\mathbb{Z}_{3}$ examples above. Of course, the mass-shell physical state 
(12.7b) is guaranteed to be null by the physical-state condition, but we have 
here computed its norm in a way that shows it is independent of the ground-state
mass-squared $P^2(\sigma)_{(0)} = 2(-1+\hat\delta_0(\sigma))$.

We will not present here the physical states at fractional levels between 
levels one and two, remarking only that there is always the familiar null physical
state at level two
\begin{subequations}
\begin{equation}
	(\alpha L(-2) + \beta L^2(-1)) | P^2(\sigma)\! =\! P^2(\sigma)_{(0)} \!+ 4 \!=\! 2(1+\hat \delta_0(\sigma))\rangle
\end{equation}
\begin{equation}
	3\alpha =2 \beta
\end{equation}
\end{subequations}
independent of the conformal-weight shifts. This is the only combination 
of mass-shell states (12.3) which is physical at level two.

All the null physical states encountered so far in this discussion are 
linear combinations of the mass-shell states (12.3), and hence removable 
in principle by the single family of gauges in Eq.~(12.4). 
Moreover the pattern of each of these examples is familiar from the special case
\begin{equation}
	D(\sigma) = 26,\quad \hat\delta_0(\sigma) = 0, \quad P^2(\sigma)_{(0)} = -2, \quad L(M) = L(M)_{(26)}
\end{equation}
which is ordinary untwisted critical open-string theory. In the following 
section, we provide further evidence for the no-ghost conjecture in the orientation-orbifold
string theories.

\section{The Ordinary $D(\sigma)\leq 26$ String Subsectors}

\noindent In this last section on the open-string sectors of the orientation-orbifolds, we discuss the \emph{ordinary} 
$D(\sigma)$-dimensional string subsystem which exists in each open-string sector $\sigma$.

This subsystem is defined by the subset of physical states $\{ | \chi[D(\sigma)] \rangle\}$ which contains \emph{none}
of the extra $(26-D(\sigma))$ fractional modes of sector $\sigma$. 
On this subset of states, the full physical-state condition (3.2a) of sector 
$\sigma$ reduces to the following description:
\begin{subequations}
\begin{equation}
	(L(M\geq 0)_{(D(\sigma))} - a(\sigma) \delta_{M,0}) |\chi[D(\sigma)]\rangle = 0
\end{equation}
\begin{equation}
	L(M)_{(D(\sigma))} = -\frac{1}{2} \eta^{ab}_{(D(\sigma))} 
		\sum_{P\in \mathbb{Z}} \normalorder J_a(P) J_b(M-P) \normalorder_M
\end{equation}
\begin{equation}
	a(\sigma)\equiv 1-\hat\delta_0(\sigma) = -\frac{1}{2} P^2(\sigma)_{(0)} \leq 1
\end{equation}
\begin{equation}
	2\leq (D(\sigma)=\text{even})\leq 26
\end{equation}
\begin{equation}
	[L(M)_{(D(\sigma))},L(N)_{(D(\sigma))}] = (M-N) L(M+N)_{(D(\sigma))} + \frac{D(\sigma)}{12} M(M^2-1) \delta_{M+N,0}
\end{equation}
\begin{equation}
	[L(M)_{(D(\sigma))},J_a(N)] = -N J_a(M+N), \quad a = 0,1,\dots,D(\sigma)-1
\end{equation}
\begin{equation}
	[J_a(M),J_b(N)] = N \eta_{ab}^{(D(\sigma))} \delta_{M+N,0}
\end{equation}
\end{subequations}
This subsector is recognized as an \emph{ordinary} $SO(D-1,1)$-invariant string
with $D=D(\sigma)\leq 26 $ space-time dimensions and a \emph{quantized intercept} 
$a(\sigma)\leq 1$, given in Eq.~(13.1c) in terms of the conformal-weight shifts.
Explicit expressions for the conformal-weight shifts and ground-state
momentum-squareds are given in Eqs.~(2.13b),(2.16) and (3.10d), leading to the explicit form
of the quantized intercept of sector $\sigma$:
\begin{subequations}
\begin{equation}
	a(\sigma) = \frac{1}{24} (d-2+\sum_L\frac{\alpha^2(L)}{F_L(\sigma)}) \leq 1
\end{equation}
\begin{equation}
	\sum_L F_L(\sigma) = 26 - d, \quad 1\leq d\leq 25.
\end{equation}
\end{subequations}
The L-cycle function $\alpha(L)$ is defined in Eq.~(2.12).
The maximal value of the quantized intercept is obtained only for $d=25,26$ where
\begin{subequations}
\begin{equation}
	D(\sigma) = 26,\quad \hat\delta_0(\sigma) = 0, \quad \hat P^2(\sigma)_{(0)} = -2,\quad a(\sigma) = 1
\end{equation}
\begin{equation}
	L(M)_{(26)} = L(M)
\end{equation}
\end{subequations}
and the full sector $\sigma$ is in these cases nothing but an ordinary critical 
open string.

Ordinary strings with $D(\sigma)\leq 26$ and $a(\sigma)\leq 1$ are known [24]
to have no negative-norm physical states, and these specifications in fact include \emph {all} of the ``ordinary'' string
subsectors discussed here (see Eq.~(13.1d)). Such consistency in the new 
string theories should not be surprising
because the twisted sectors of these theories are constructed from (copies of) critical untwisted theories using
only  the principles of orbifold theory.

It should be mentioned that each of these ``ordinary'' open-string subsectors
by themselves would give rise to continuous closed-string spectra via 
non-planar loops. In our theories however, each full open-string sector $\sigma$
also contains the extra $(26-D(\sigma))$ fractional-moded Lorentz-scalar degrees of freedom. 
Since the total number of effective degrees of freedom of each sector is 
26, we do not expect any such difficulty in the loops. This intuition is 
supported by the closed-string spectra of our orbifold construction in the 
following sections.

Motivated in part by our interest in null-physical states (see Sec.~12), we also give here
the explicit solution of Eq.~(13.1a) for the low-lying physical states $\{ | \chi[D(\sigma)] \}$ in the ordinary string subsectors.

\noindent \underline{level 0}

Here the only physical state is the familiar positive-norm oscillator-free state
\begin{equation}
	| P^2(\sigma) \!=\! P^2(\sigma)_{(0)} \!=\! - 2 a(\sigma) \rangle, \quad (J=0 \text{ ground state})
\end{equation}
which is also the true ground-state of each full sector $\sigma$.
\newpage
\noindent \underline{level 1}

Here the only physical state for $D(\sigma)\leq 25$ is the familiar $J=1$ state
\begin{subequations}
\begin{equation}
	\epsilon(\sigma)\!\cdot\! J(-1) | P^2(\sigma)\! =\! P^2(\sigma)_{(0)} \!+ 2 \rangle, \quad \epsilon(\sigma)\!\cdot\! P(\sigma) = 0, \quad \epsilon^2(\sigma) = -1
\end{equation}
\begin{equation}
	||\,\, | \dots \rangle ||^2 = || \,\,|P^2(\sigma) \!=\! P^2(\sigma)_{(0)}\! + 2 \rangle ||^2 
\end{equation}
\end{subequations}
which is a massive vector meson with positive norm. 

The $J=0$ state
\begin{subequations}
\begin{equation}
	L(-1)_{(D(\sigma))} | P^2(\sigma)\! =\! P^2(\sigma)_{(0)}\! + 2\rangle = -P(\sigma)\!\cdot\! J(-1) |P^2(\sigma)\! =\! P^2(\sigma)_{(0)}\! + 2\rangle
\end{equation}
\begin{multline}
	|| L(-1)_{(D(\sigma))} |P^2(\sigma)\! =\! P^2(\sigma)_{(0)} \!+ 2\rangle ||^2\\
		= \langle P^2(\sigma)\! = \!P^2(\sigma)_{(0)} \!+ 2 |\, 2 L(0)_{(D(\sigma))} | P^2(\sigma)\! =\! P^2(\sigma)_{(0)} \!+ 2\rangle
\end{multline}
\begin{equation}
 	= -P^2(\sigma) || \,|P^2(\sigma)\! = \!P^2(\sigma)_{(0)} \!+ 2\rangle ||^2
\end{equation}
\end{subequations}
is \emph{not} physical and \emph{not} null, except of course at
\begin{subequations}
\begin{equation}
	D(\sigma) = 26, \quad L(M) = L(M)_{(26)}
\end{equation}
\begin{equation}
	a(\sigma) = 1,\quad P^2(\sigma)_{(0)} = -2,\quad P^2(\sigma) = 0
\end{equation}
\end{subequations}
where the entire sector $\sigma$ of the orientation-orbifold is an ordinary critical open string.

It is clear that the state (13.6) for $D(\sigma)\leq 25$ is \emph{not}
the null-physical state (10.6c) discussed above
\begin{subequations}
\begin{equation}
	L(-1) |P^2(\sigma)\! = \!P^2(\sigma)_{(0)} \!+2 \rangle = (L(-1)_{(D(\sigma))} + \text{extra}) |P^2(\sigma)\! = \!P^2(\sigma)_{(0)} \!+ 2\rangle 
\end{equation}
\begin{equation}
	|| L(-1) |P^2(\sigma)\! =\! P^2(\sigma)_{(0)} \!+2 \rangle ||^2 = 0
\end{equation}
\end{subequations}
which invoves the \emph{full} Virasoro generators $\{ L(M) \}$, and which should
be removed by the single set of gauges (12.4) associated to the full 
Virasoro generators. Put another way, we expect that \emph {no extra set of gauges}
based on the subsector generators $\{ L(M)_{(D(\sigma))} \}$ will be 
necessary to remove all the null-physical states.

\noindent \underline{level 2}

We find the familiar $J=2$ physical state
\begin{equation}
	\{ \epsilon^a(\sigma) \epsilon^b(\sigma) + \frac{1}{D(\sigma)-1} (\eta^{ab}_{(D(\sigma))} -
	 \frac{P^a(\sigma) P^b(\sigma)}{P^2(\sigma)} J_a(-1) J_b(-1)\} |P^2(\sigma) \!=\! P^2(\sigma)_{(0)} \!+ 4\rangle
\end{equation}
with positive norm for all $D(\sigma)\leq 26$. There is also a $J=1$ physical state
\begin{subequations}
\begin{equation}
	\{ 2 \epsilon(\sigma) \!\cdot \!J(-1) P(\sigma) \!\cdot\! J(-1) - P^2(\sigma) \epsilon(\sigma)\!\cdot\! J(-2)\} |P^2(\sigma)\! =\! P^2(\sigma)_{(0)} \!+ 4\rangle
\end{equation}
\begin{equation}
	|| \, |\dots\rangle ||^2 = 2(P^2(\sigma)_{(0)} + 4)(P^2(\sigma)_{(0)} + 2) \,||\,\, |P^2(\sigma) \!=\! P^2(\sigma)_{(0)}\! + 4\rangle ||^2
\end{equation}
\end{subequations}
whose norm is positive except for ground-state mass-squared $P^2(\sigma)_{(0)} = -2$, where the vanishing norm marks again an ordinary $D(\sigma)=26$ dimensional
open string.

To supplement the discussion of the state (13.10), consider also the following $J=1$ state:
\begin{subequations}
\begin{multline}
	{L(-1)}_{(D(\sigma))} \epsilon(\sigma) \!\cdot\! J(-1) | P^2(\sigma)\! = \! P^2(\sigma)_{(0)} \!+ 4\rangle  \\
		= \{ \epsilon(\sigma)\!\cdot\! J(-2) - \epsilon(\sigma)\!\cdot\! J(-1) P(\sigma)\!\cdot\! J(-1) \} |P^2(\sigma) \!=\! P^2(\sigma)_{(0)} \!+ 4\rangle
\end{multline}
\begin{equation}
	||\, |\dots \rangle ||^2 = (P^2(\sigma)_{(0)} + 2)\,\,\, ||\,\, |P^2(\sigma) = P^2(\sigma)_{(0)} + 4\rangle ||^2
\end{equation}
\end{subequations}
This state is neither physical nor null -- except at $P^2(\sigma)_{(0)} = -2$ and 
hence $D(\sigma) =26$. At 
such a point the full sector $\sigma$ is again an ordinary critical 
string, and this state -- now proportional to the state (13.10a)) --
is a familiar null-physical state which is removed by the single set of 
gauges (12.4) with full Virasoro generators $L(M)=L(M)_{(26)}$. 
Once again, consistent with our conjecture, no extra family of gauges based on 
the subsector generators $\{ L(M)_{(D(\sigma))} \}$ is required.

Finally, level two contains a single $J=0$ physical state:
\begin{subequations}
\begin{equation}
	\{ P(\sigma)\!\cdot \!J(-2) - \frac{(D(\sigma) + 2P^2(\sigma))}{P^2(\sigma)(D(\sigma)-1)} (P(\sigma)\!\cdot\! J(-1))^2 + 
		\frac{(2P^2(\sigma)+1)}{D(\sigma)-1} J(-1)\!\cdot\! J(-1) \} | P^2(\sigma) \!=\! P^2(\sigma)_{(0)} \!+ 4\rangle
\end{equation}
\begin{equation}
	||\,\, |\dots \rangle ||^2 = 2\{ 1-P^2(\sigma) + \frac{(2P^2(\sigma)+1)^2}{D(\sigma)-1} \} ||\,\, |P^2(\sigma)\! 
	=\! P^2(\sigma)_{(0)} \!+ 4\rangle ||^2.
\end{equation}
\end{subequations}
The norm of the $J=0$ physical state is positive for $P^2(\sigma)$ 
outside the region between the zeroes of the norm, which are located at 
the values:
\begin{equation}
	P_{\pm}^2(\sigma) = \frac{1}{8} ( D(\sigma) - 5\pm \sqrt{(D(\sigma)-1)(D(\sigma)-25)}) = \text{real}.
\end{equation}
It follows that the norm of the $J=0$ physical state is strictly positive for 
\begin{equation}
	||\,\, |\dots\rangle ||^2 > 0 : \quad 2\leq D(\sigma) \leq 24
\end{equation}
because the zeros are complex in this range.

Beyond this range, we consider three additional cases.\\
\noindent a) $D(\sigma)=25$. Here the zeroes of the norm coincide
at $P_\pm^2(\sigma)= 5/2$, and we find strictly positive norm except zero norm at
$P^2(\sigma)=5/2$, $P^2(\sigma)_{(0)} = -3/2$ where the $J=0$ state is proportional to
\begin{equation}
	-\frac{1}{2} (L(-2)_{(25)} + L^2(-1)_{(25)} ) |P^2(\sigma)\!=\! \frac{5}{2}\rangle.
\end{equation}
\noindent In ordinary $D=25$ dimensional string theory, this 
null-physical state would be removed by a set of gauges constructed from the Virasoro generators 
$\{ L(M)_{(25)}\}$, but for us this would be an \emph {extra} set of 
gauges , beyond those constructed in (12.4) with the total Virasoro generators $\{L(M) = L(M)_{(25)} + (\text{extra}) \}$.
Fortunately, the space-times  of the open-string sectors of the 
orientation-orbifold string theories  satisfy the double-inequality of Sec.~7
\begin{equation}
	2\leq D(\sigma) = \text{even} \leq 26
\end{equation}
and therefore contain \emph {no such sectors with $D(\sigma)= \text{odd}$}. We consider this observation 
as a non-trivial check in support of our conjecture that the single set 
of gauges (12.4) is sufficient to remove all null-physical states in the 
new string theories.

\noindent b) $D(\sigma)=26$. Here we have

\begin{equation}
	P_+^2(\sigma) = \frac{13}{4},\quad P^2_-(\sigma) = 2,\quad P^2(\sigma)_{(0)} = -2,\quad P^2(\sigma)=2.
\end{equation}
This is the familiar case of the critical open string, where the $J=0$ physical state (13.12 ) is null,
proportional to
\begin{subequations}
\begin{equation}
	-\frac{1}{5} (2L(-2)_{(26)} + 3L^2(-1)_{(26)} ) | P^2(\sigma)\! =\! 2\rangle
\end{equation}
\begin{equation}
	L(M)_{(26)} = L(M)
\end{equation}
\end{subequations}
and removed by the single set of gauges (12.4).

\noindent c) $D(\sigma)\geq 27$. Above $D(\sigma)=26$, it is well-known [24]
that the $J=0$ physical state (13.22) can have negative norm in the range $P_-^2(\sigma)\leq P^2(\sigma) \leq P^2_+(\sigma)$.
Fortunately again, the open-string sectors of the orientation-orbifold 
string systems contain no such high-dimensional space-times.

Consistent with the low-level examples of this section, we finally emphasize 
the following. If our 
single-family gauge conjecture in Sec.~12 is correct then the ordinary-string subsectors
(13.1) will contain \emph{no null-physical states at all} except at 
$D(\sigma)=26$ (the critical open string), where our single set of gauges with  $\{ 
L(M) = L(M)_{(26)}\}$ is of course sufficient to remove the null-physical states.

\section{Assembling the Orientation-Orbifolds}

\noindent We recall here that the general orientation-orbifold string 
theory [20,21,23,1,3,4,8] has the form
\begin{equation}
  \frac{U(1)^{26}}{H_-} = \frac{U(1)^{26}_L \times U(1)^{26}_R}{H_-}, \quad H_- \subset \mathbb{Z}_2(w.s.) \times H'_{26}
\end{equation}
where $\mathbb{Z}_2(w.s.) = (\tau_0 = 1, \tau_-)$ is the group of 
left-right exchanges on the closed string $U(1)^{26}$,
and the space-time symmetry group $H'_{26}$ acts on the left- and right-movers 
separately. The elements of the group $H_-$ associated to $\tau_-$ correspond 
to twisted open-string sectors at $\hat c(\sigma) = 52$ (and reduced central charge 
$c(\sigma) = 26$). A large subset of these open-string sectors have been studied at length above,
and earlier in Refs.~[3,4,8].

The orientation-orbifold string systems always contain an equal number of 
twisted open- and closed-string sectors, the latter corresponding to those
elements of $H_-$ proportional to
the trivial element $\tau_0=1$ of $\mathbb{Z}_2(w.s.)$. The closed-string sectors
therefore comprise the ordinary space-time orbifold
\begin{equation}
  \frac{U(1)^{26}}{H'_{26}}
\end{equation}
all of whose sectors live at central charge $\hat c(\sigma) = 26$. Note that we have 
chosen here to suppress the element  $\tau_0=1$, whose action is trivial on the
closed string $U(1)^{26}$.

More precisely, we focus our attention on the large example (1.5)
\begin{subequations}
\begin{equation}
	H_- \subset  \{ H'_{26};\,\, \tau_- \times (\pm \thickone)_{(d)} \times H(\text{perm})'_{26-d} \}
\end{equation}
\begin{equation}
	H'_{26}\subset (\thickone)_{(d)} \times H(\text{perm})'_{26-d}
\end{equation}
\end{subequations}
where the closed- and open- string sectors appear 
respectively before and after the semicolon. Recall from our discussion 
above that we may choose either  $(\pm\thickone)_{(d)}$ that is 
$\epsilon= 0\text{ or } 1$ (but not both) for these open-string sectors, 
and moreover that
the physical spectrum of each open-string sector is Lorentzian and independent of $\epsilon$.
On the other hand, we are choosing only  $\epsilon =0$ for the closed-string sectors,
which then guarantees [8] that all the closed-string sectors in the large 
example (14.3) are also Lorentzian.
Following our practice for the open-string sectors,  we will again choose the cyclic groups
\begin{equation}
	H(\text{perm})'_{26-d} = \mathbb{Z}_{26-d}
\end{equation}
when needed as explicit examples in the closed-string sectors. 

The reader should  bear in mind that there is in general more than 
one way to choose 
the groups $H_-$ and $H'_{26}$ so that $H'_{26}\subset H_-$. As examples,
consider the cyclic groups (14.4), where we may choose for any $d$ (see Eq.~(7.1)):
\begin{subequations}
\begin{equation}
	H_-=\{ H'_{26};\,\, \tau_- \times (\pm\thickone)_{(d)} \times (\thickone, \omega, \dots, \omega^{25-d})_{26-d} \}
\end{equation}
\begin{equation}
	H'_{26}= (\thickone)_{(d)} \times (\thickone, \omega, \dots, \omega^{25-d} )_{26-d}
\end{equation}
\begin{equation}
	\omega \in \mathbb{Z}_{26-d}.
\end{equation}
\end{subequations}
Then the space-time orbifold $U(1)^{26}/H_{26}'$ in Eq. (14.2) is an 
ordinary cyclic permutation orbifold [9] on the $(26-d)$ 
 higher spatial dimensions of $U(1)^{26}$. On the other
hand, when $d$ is even we may alternately choose the restricted groups
\begin{subequations}
\begin{equation}
	H_-=\{ H'_{26};\,\, \tau_- \times (\pm\thickone)_{(d)} \times (\omega, \omega^3,\dots, \omega^{25-d})_{26-d} \}
\end{equation}
\begin{equation}
	H'_{26}= (\thickone)_{(d)} \times (\thickone,\omega^2, \dots, \omega^{24-d})_{26-d} 
\end{equation}
\end{subequations}
where $H'_{26}$ is now proportional to  the subgroup $\mathbb{Z}_{(26-d)/2}$ of $\mathbb{Z}_{26-d}$.
Other choices of $H'_{26}\subset H_-$ are possible for special values of 
$d$, but we will not pursue this further here.

\section{The Twisted Closed-String Sectors}

\noindent Using the standard methods of the orbifold program [10-23], we have 
worked out the physical-state condition and Virasoro generators for each closed-string sector $\sigma$ corresponding to 
any $\omega'(\sigma) \subset H'_{26}$ of the ordinary space-time orbifold 
$U(1)^{26}/H_{26}'$. The results for general $ H'_{26}$ are as follows:
\begin{subequations}
  \begin{equation}
    (\hat L(m\geq 0)-\delta_{m,0})|\chi(\sigma)\rangle = 
    (\bar{\hat L}(m\geq 0)-\delta_{m,0}) |\chi(\sigma)\rangle = 0
  \end{equation}
  \begin{multline}
    \hat L(m) = \frac{1}{2} \mathcal{G}^{n(r)\mu ; - n(r),\nu} (\sigma)
    \sum_{p\in\mathbb{Z}}\normalorder \hat J_{n(r)\mu}(p+\frac{n(r)}{\rho(\sigma)}) \hat J_{-n(r),\nu} (m-p-\frac{n(r)}{\rho(\sigma)})\normalorder_M + \\
      + \delta_{m,0} \frac{1}{4} \sum_r \text{dim}[\bar n(r)] \frac{\bar n(r)}{\rho(\sigma)}(1-\frac{\bar n(r)}{\rho(\sigma)})
  \end{multline}
  \begin{equation}
    \sum_{r\mu} = \sum_r \text{dim}[\bar n(r)] = 26
  \end{equation}
  \begin{equation}
    [\hat L(m), \hat L(n)] = (m-n)\hat L(m+n) + \frac{26}{12} m(m^2-1) \delta_{m+n,0}
  \end{equation}
  \begin{equation}
    [\hat L(m), \hat J_{n(r)\mu}(n+\frac{n(r)}{\rho(\sigma)})] = 
    -(n+\frac{n(r)}{\rho(\sigma)}) \hat J_{n(r)\mu}(m+n+\frac{n(r)}{\rho(\sigma)})
  \end{equation}
  \begin{equation}
    [\hat J_{n(r)\mu} (m+\frac{n(r)}{\rho(\sigma)}), \hat J_{n(s)\nu} (n+\frac{n(s)}{\rho(\sigma)})] = 
    (m+\frac{n(r)}{\rho(\sigma)}) 
    \delta_{m+n+\frac{n(r)+n(s)}{\rho(\sigma)},0} \mathcal{G}_{n(r)\mu ; -n(r),\nu}(\sigma)
  \end{equation}
  \begin{equation}
    \hat J_{n(r)\pm \rho(\sigma),\mu}(m + \frac{n(r)\pm \rho(\sigma)}{\rho(\sigma)}) = \hat J_{n(r)\mu}(m\pm 1 + \frac{n(r)}{\rho(\sigma)}).
  \end{equation}
\end{subequations}
There is also a set of right-mover Virasoro generators $\{ \bar{\hat 
L}\}$  which are copies (with right mover currents $\{\bar{\hat J}\}$) of 
the left-mover system given explicitly here.

For these closed-string sectors, the physical-state conditions (15.1a) 
are quite ordinary, and the twisted metric $\mathcal{G}(\sigma)$ in these results
has the usual form [11,13,3,7]:
\begin{subequations}
	\begin{equation}
		\mathcal{G}_{n(r)\mu; n(s)\nu}(\sigma) = \chi_{n(r)\mu}(\sigma) \chi_{n(s)\nu}(\sigma) {U(\sigma)_{n(r)\mu}}^a {U(\sigma)_{n(s)\nu}}^b G_{ab}
	\end{equation}
	\begin{equation}
		= \delta_{n(r)+n(s),0\text{ mod } \rho(\sigma)} \mathcal{G}_{n(r)\mu;-n(r),\nu}(\sigma)
	\end{equation}
	\begin{equation}
		{\omega'(\sigma)_a}^b {U^\dagger(\sigma)_b}^{n(r)\mu} = e^{-2\pi i \frac{n(r)}{\rho(\sigma)}} {U^\dagger(\sigma)_a}^{n(r)\mu}, \quad \omega'(\sigma) \in H'_{26}
	\end{equation}
	\begin{equation}
		G_{ab} = -\eta_{ab}, \quad \eta = \left( 
			\begin{array}{cc}
				1 & 0 \\
				0 & -\thickone
			\end{array}
		\right), \quad a,b = 0,1,\dots,25.
	\end{equation}
\end{subequations}
Here $\{n(r)\}$ and $\rho(\sigma)$ are respectively the spectral indices and the order 
of $\omega'(\sigma)$, and the mode normal-ordering follows the standard 
convention (2.4).

The twisted closed-string currents above exhibit the conventional orbifold fractions ($n/\rho$) at $\hat c(\sigma) = 26$,
and there is no distinct reduced formulation for these cases. These results can be 
understood as cases of trivial cycle length $f_{j}(\sigma)=1$ in the more general reduced orbifold fraction  
$f_j(\sigma)n/\rho$ [7,8] at $c(\sigma) = 26$.

Next, we give the explicit form of the  closed-string Virasoro 
generators  in the case of the large example (14.3), where the space-time
orbifold has the form
\begin{equation}
	\frac{U(1)^{26}}{H(\text{perm})'_{26-d}}.
\end{equation}
Including now the adjoint operations [13,8], these results are as follows:
\begin{subequations}
\begin{multline}
	\hat L(m) = \delta_{m,0} \frac{1}{24} \sum_L(F_L(\sigma) - \frac{1}{F_L(\sigma)}) - \frac{1}{2} \eta^{ab}_{(D(\sigma)_c)} \sum_{p\in\mathbb{Z}} 
		\normalorder J_a(p) J_b(m - p) \normalorder_M + \\
		+ \frac{1}{2} \sum_L \frac{1}{F_L(\sigma)} \sum_{\hat L=1}^{F_L(\sigma)-1} \sum_{p\in\mathbb{Z}} \normalorder \hat J_{\hat L L}(p + \frac{\hat L}{F_L(\sigma)} ) \hat J_{-\hat L,L} (m - p - \frac{\hat L}{F_L(\sigma)})\normalorder_M
\end{multline}
\begin{equation}
	[\hat L(m),\hat L(n)] = (m-n) \hat L(m+n) + \frac{26}{12} m(m^2-1) \delta_{m+n,0}
\end{equation}\begin{equation}
	[\hat L(m), J_a(n)] = -n J_a(m+n), \quad a=0,1,\dots, D(\sigma)_c -1
\end{equation}\begin{equation}
	[\hat L(m), J_{\hat LL}(n + \frac{\hat L}{F_L(\sigma)})] = - (n+\frac{\hat L}{F_L(\sigma)}) \hat J_{\hat LL}(m+n+\frac{\hat L}{F_L(\sigma)})
\end{equation}\begin{equation}
	\hat L=1,\dots, F_L(\sigma)-1
\end{equation}\begin{equation}
	[\hat J_a(m), \hat J_b(n)] = n \eta^{(D(\sigma)_c)}_{ab} \delta_{m+n,0}
\end{equation}\begin{equation}
	[\hat J_{\hat LL}(m+\frac{\hat L}{F_L(\sigma)}), \hat J_{\hat MM}(n+\frac{\hat M}{F_M(\sigma)})] = 
		\delta_{LM} (m+\frac{\hat L}{F_L(\sigma)}) F_L(\sigma) \delta_{m+n+\frac{\hat L+\hat M}{F_L(\sigma)}, 0}
\end{equation}\begin{equation}
	\hat J_{\hat L\pm F_L(\sigma), L}(m + \frac{\hat L\pm F_L(\sigma)}{F_L(\sigma)}) = \hat J_{\hat LL}(m\pm 1 + \frac{\hat L}{F_L(\sigma)})
\end{equation}\begin{equation}
	J_a(m)^\dagger = J_a(-m)
\end{equation}\begin{equation}
	\hat J_{\hat LL}(m + \frac{\hat L}{F_L(\sigma)})^\dagger = \hat J_{-\hat L,L}(-m - \frac{\hat L}{F_L(\sigma)})
\end{equation}\begin{equation}
    \hat{L}(m)^\dagger = \hat{L}(-m)	
\end{equation}\begin{equation}
	\sum_L F_L(\sigma) = 26-d, \quad \sum_a + \sum_{L,\hat L\neq 0} =26, \quad \sum_L = N(\sigma)'.
\end{equation}
\end{subequations}
Here, as above for the twisted open-string sectors, $F_L(\sigma)$ is the length of cycle $L$ in $\omega'(\sigma) \in H(\text{perm})'_{26-d}$
and $N(\sigma)'$ is the total number of cycles in $\omega'(\sigma)$. The 
closed-string target space-time dimensionality $D(\sigma)_c$ is evaluated 
in the next paragraph.

Following our development of the open-string sectors above, we have here directly collected
\emph{all} the integer-moded sequences in the second term of the Virasoro 
generators (15.4a). This includes in particular the $\hat L=0$ terms for each cycle $L$ in $\omega'(\sigma)$.
(See the rescaling for $\{J_{0L}(M)\}$ in Eq.~(4.4a); there is no 
analogue of the integer-moded sequence 
$\{J_{F_{L}(\sigma)/2,L}(M)\}$ in the closed-string sectors.)
Thus we see that closed-string sector $\sigma$ exhibits the enhanced Lorentz symmetry $SO(D(\sigma)_c-1,1)$, 
where the number of closed-string target space-time dimensions $D(\sigma)_c$ is 
\begin{subequations}
\begin{equation}
	2\leq D(\sigma)_c = d + N(\sigma)' = d+ N_O(\sigma)' + N_E(\sigma)' \leq 26
\end{equation}
\begin{equation}
	2\leq D(\sigma) = d + N_O(\sigma)' + 2 N_E(\sigma)' = \text{even} \leq 26.
\end{equation}
\end{subequations}
The quantities $N_{O,E}(\sigma)'$ are again the number of cycles of odd and even length $F_L(\sigma)$ in $\omega'(\sigma)$.
In Eq.~(15.5b) we have also provided for comparison the number (7.6) of 
target space-time dimensions in the corresponding open-string sector 
$\sigma$. (We remind that both the open- and closed-string space-time dimensions are associated to 
the same element $\omega'(\sigma)\in H(\text{perm})'_{26-d}$ as 
$\tau_0\times\omega'(\sigma)$ and $\tau_-\times\omega'(\sigma)$ 
respectively.) 

Notice that the following triple inequality   
\begin{equation}
   2 \leq D(\sigma)_{c} \leq D(\sigma) \leq 26
\end{equation}
follows for all $H(\text{perm})'_{26-d}$ by comparison of the explicit formula (15.5a) with our 
previous result (15.5b). Moreover, the equality 
$D(\sigma)_{c}=D(\sigma)$ is achieved only 
when all cycles of $\omega'(\sigma)$ have odd length, and we remark in 
particular that the space-time dimensionalities of the closed strings are not necessarily
even.

The last two columns of Tables 1, 2 and 3 in Sec.~7 give the explicit values
of the closed-string target-space dimensionalities  for the non-trivial elements of the cyclic groups $H(\text{perm})'_{26-d} = \mathbb{Z}_{26-d}$. 
We shall return below to some special non-trivial cases in these tables, 
mentioning here only that the trivial closed-string sector $\sigma=0$ 
with $\omega'(0)=1$ is an ordinary untwisted $D(0)_{c}=26$ dimensional closed 
string for all $H(\text{perm})'_{26-d}$.

For completeness, we have used the usual closed-string conditions $\hat J(0)^L = \hat J(0)^R \equiv \hat J(0)/\sqrt{2}$
to find the ground-state mass-squared of each closed-string sector $\sigma$ of 
the full space-time orbifold in Eq.~(15.3):
\begin{subequations}
\begin{equation}
P^2(\sigma)^c_{(0)} = -\frac{1}{6} \{ (d-2) + \sum_L \frac{1}{F_L(\sigma)} \} \geq -4
\end{equation}
\begin{equation}
	F_L(\sigma) \geq 1, \quad \sum_L F_L(\sigma) = 26-d, \quad 1\leq d\leq 25.
\end{equation}
\end{subequations}
This result for closed-string sector $\sigma$ should be compared with the corresponding
ground-state mass-squared of open-string sector $\sigma$ in Eq.~(3.10d). As for 
the open-string sectors, this result can be extended to $d=26$ by 
ignoring the summation over $L$, and we conclude that the extrema of 
Eqs.~(15.5a) and (15.7a) are obtained only in the two cases with trivial $H(\text{perm})'$
\begin{equation}
	d=25,26: \quad D(0)_c=26, \quad P^2(0)^c_{(0)} = -4
\end{equation}
where the only sector $\sigma=0$ is again an ordinary untwisted  critical closed string. 

It is not difficult to check from Eq.~(15.7a) that all the sectors of the 
space-time orbifold (15.3) are tachyonic, except possibly for the case $H(\text{perm})'_{25}$ where 
\begin{equation}
	d=1: \quad P^2(\sigma)^c_{(0)} = \frac{1}{6} (1-\sum_L\frac{1}{F_L(\sigma)}),\quad \sum_L F_L(\sigma)=25.
\end{equation}
Indeed one sees explicitly from the cycle-data in Table 2 that all non-trivial sectors of $\mathbb{Z}_{25}$ are non-tachyonic
\begin{subequations}
\begin{equation}
	20[25]^1: \quad D(\sigma)_c = 2,\quad P^2(\sigma)^c_{(0)} = \frac{4}{25}
\end{equation}
\begin{equation}
	4[5]^5:\quad D(\sigma)_c = 6,\quad P^2(\sigma)^c_{(0)} = 0
\end{equation}
\end{subequations}
with enhanced Lorentz symmetry $SO(1,1)$ and $SO(5,1)$ respectively.

As a simple subclass of examples, consider the closed- and open-string 
sectors corresponding to the non-trivial elements of the prime cyclic groups.
Here we find $(25-d)$ non-trivial single-cycle closed- and open-string sectors with
\begin{subequations}
\begin{equation}
	H(\text{perm})'_{26-d} = \mathbb{Z}_{26-d}, \quad d = \text{prime}, \quad \sigma = 1,\dots, 25-d
\end{equation}
\begin{equation}
	D(\sigma)_c = D(\sigma) = d+1
\end{equation}
\begin{equation}
	P^2(\sigma)^c_{(0)} = 2P^2(\sigma)_{(0)} = -\frac{1}{6}(d-2 + \frac{1}{26-d})
\end{equation}
\end{subequations}
where the open-string sectors were given above in Eq.~(6.3). 

Among these closed-string sectors, we mention in particular the $22$ non-trivial closed-string sectors of $H(\text{perm})'_{23} = \mathbb{Z}_{23}$, 
which are \emph{four-dimensional} closed strings with enhanced Lorentz symmetry $SO(3,1)$:
\begin{subequations}
\begin{multline}
	\hat L(m) = \frac{22}{23} \delta_{m,0} - \frac{1}{2} \eta^{ab}_{(4)} \sum_{p\in\mathbb{Z}}
		\normalorder \hat J_a(p) \hat J_b(m-p)\normalorder_M +\\
		+ \frac{1}{46} \sum_{\hat L=1}^{22} \sum_{p\in\mathbb{Z}} \normalorder 
			\hat J_{\hat L0}(p + \frac{\hat L}{23}) \hat J_{-\hat L,0}(m-p-\frac{\hat L}{23})\normalorder_M
\end{multline}
\begin{equation}
	d=3, \quad D(\sigma)_c=4,\quad P^2(\sigma)^c_{(0)} = 2 P^2(\sigma)_{(0)} = -\frac{4}{23},\quad \sigma = 1, \dots, 22.
\end{equation}
\end{subequations}
In fact, the left-mover Virasoro generators of these 22 closed 
four-dimensional strings are isomorphic to the 22 four-dimensional 
open-string sectors given for $\mathbb{Z}_{23}$ in Eq.~(9.1). This conclusion 
follows by mode-relabeling under the mode sums, so that
$\{ 2\hat L/F_L(\sigma)\} \sim \{ \hat L/F_L(\sigma)\}$ when $F_L(\sigma)$
is odd (see the conclusions of Ref.~[3]).

Surveying the Tables of the cyclic groups in Sec.~7, we find that the only
other four-dimensional $SO(3,1)$-invariant closed string is the single sector $1[12]^2$ of $H(\text{perm})'_{24} = \mathbb{Z}_{24}$.

We conclude this section with some general remarks about the full 
orientation-orbifold string systems (14.3), including both the open- and 
closed-string sectors, first for the cyclic groups $H(\text{perm})'_{26-d} = \mathbb{Z}_{26-d}$ as
described by the general sector schematics in Eq.~(7.2). One universal feature 
of all these orbifolds is the presence of the \emph {ordinary} critical $c=26$ 
open-closed string systems (see Ref.~[4]) as the following two sectors
\begin{subequations}
\begin{equation}
	\tau_-\times (\pm\thickone)_{(d)} \times (\thickone)_{(26-d)},\quad\quad \tau_0\times (\thickone)_{(d)} \times (\thickone)_{(26-d)}
\end{equation}
\begin{equation}
	(\thickone)_{(26-d)}\in\mathbb{Z}_{26-d}, \quad 1\leq d\leq 26
\end{equation}
\begin{equation}
	D(\sigma) = D(\sigma)_c = 26,\quad P^2(\sigma)^c_{(0)} = 2P^2(\sigma)_{(0)} = -4
\end{equation}
\end{subequations}
both of which correspond to the trivial element of $\mathbb{Z}_{26-d}$. 
We remind that the physical spectrum of the open-string sector $\tau_{-}$ is 
independent of the choice $(\pm\thickone)_{(d)}$, and
for the cases $d=25,26$ these two sectors form the entire 
orientation-orbifold. 
Similarly, the ordinary critical open-closed string subsystem (15.13)  persists  for the trivial elements $ (\thickone)_{(26-d)}$
of all $H(\text{perm})'_{26-d}$.

Moreover, for all $H(\text{perm})'_{26-d}$, the closed-string sector
\begin{equation}
	\thickone = \tau_0\times (\thickone)_{(d)} \times (\thickone)_{(26-d)}
\end{equation}
is the \emph{only} ordinary closed-string sector, so that each 
orientation-orbifold string system (14.3) contains exactly \emph{one graviton per orbifold} --
in keeping with our conventional understanding of gravity in string theory.

It should be emphasized however that the orientation-orbifold string 
systems 
are uniquely simple in this regard. In contrast, the generalized permutation orbifolds [8]
\begin{equation}
 \frac{U(1)^{26K}}{H_+},\quad H_+\subset H(\text{perm})_K\times (\pm \thickone)_{(d)} \times H(\text{perm})'_{26-d}
\end{equation}
are composed entirely of closed-string sectors, and can exhibit 
\emph{multiple, presumably non-interacting gravitons} (see Ref.~5 and Sec.~10 of Ref.~7). 

\section{Conclusions and Directions}

\noindent We have studied the following large example of orientation-orbifold string 
theories [1-4,6-8]
\begin{subequations}
\begin{equation}
	\frac{U(1)^{26}}{H_-} = \frac{U(1)^{26}_L \times U(1)^{26}_R}{H_-}
\end{equation}
\begin{equation}
	H_- \subset  \{ H'_{26};\,\, \tau_- \times (\pm \thickone)_{(d)} \times H(\text{perm})'_{26-d} \}
\end{equation}
\begin{equation}
	H'_{26}\subset (\thickone)_{(d)} \times H(\text{perm})'_{26-d}, \quad  1\leq d\leq 26
\end{equation}
\end{subequations}
in some detail, emphasizing in particular the cyclic permutation groups $H(\text{perm})'_{26-d}=\mathbb{Z}_{26-d}$.
These new string theories provide multi-sector generalizations of 
ordinary critical open-closed string theory, each system containing an equal number of twisted open- and 
closed-string sectors. Moreover, each system includes the ordinary critical 
open-closed string theory as the special subsystem with the unique graviton 
of the theory. The open-string sectors (see Secs.~2-7) have a local description [1,6] at 
$\hat{c}=52$  and an equivalent, reduced description [3-5,7,8] of the 
physical states at $c=26$, while the 
closed strings (see Secs.~14 and 15) form an ordinary space-time orbifold with all sectors at $\hat{c}=c=26$.

We have found that the open- and closed-string sectors of these theories 
have sector-dependent Lorentzian [8]
target space-time dimensions
\begin{equation}
	D(\sigma), D(\sigma)_{c}\leq 26
\end{equation}
for all sectors $\sigma$ associated to each $H(\text{perm})'_{26-d}$, including the corresponding Lorentz symmetries $SO(D(\sigma)-1,1)$ and
$SO(D(\sigma)_c-1,1)$. The enhancement mechanism [8] which underlies these 
symmetries is detailed in Sec.~4. See also the discussion of Sec.~7, and in particular the 
Tables given there for the cyclic 
groups -- as well as the general results in Eqs.~(15.5-6). We emphasize with Refs.~[3,8] that 
these constructions are generically new, and certainly not ordinary 
compactifications. Additionally, we have here pointed out those special cases 
of non-tachyonic or four-dimensional strings (see Secs.~8,9 and 15) associated to the cyclic groups. 

We have also included a number of introductory, successful tests of the 
no-ghost conjecture [1-5] for these theories (see Secs.~10-13).

It seems that the orientation-orbifold string systems studied here are the 
simplest among the orbifold-string theories of permutation-type [1-8], not 
least because they exhibit only a single graviton per orbifold and contain
ordinary ghost-free $D(\sigma)\leq26$ - dimensional string subsystems with quantized
intercept $a(\sigma)\leq 1$ (see Sec.~13). A next step in this program should be the study of
the twisted $\hat{c}=52$ open-string vertex operators of the orientation-orbifolds and 
the construction of the open-string sectors at tree level, following the text 
and Appendix of Ref.~4. Given our historical understanding [24], one expects
that these sectors will be the simplest in which to elevate the no-ghost
conjecture to a theorem for the new string theories. 

We conclude this paper with a final remark on another important
direction in the program. On the basis of the
enhanced target-space Lorentz symmetries [8] studied here for the bosonic prototypes,
we expect correspondingly-enhanced target-space supersymmetries -- and 
non-tachyonic spectra -- in the 
superstring generalizations [1] of the orientation-orbifold string systems. 
\newpage

\section*{Acknowledgements}

\noindent For helpful discussion and encouragement, I thank L. Alvarez-Gaum\'e, C. 
Bachas, J. de Boer, S. Frolov, O. Ganor, E. Kiritsis, A. Neveu, H. Nicolai,
N. Obers, B. Pioline, M. Porrati, E. Rabinovici, V. Schomerus, C. Schweigert,
M. Staudacher, R. Stora, C. Thorn, E. Verlinde and J.-B. Zuber.

\end{document}